\documentclass[oldversion]{aa}
\usepackage[pdftex]{graphicx}
\usepackage{natbib}
\usepackage{txfonts}
\bibpunct{(}{)}{;}{a}{}{,}

\def\z{\phantom{0}}

\def\cc{3C~273}
\def\nh{$N_{\mathrm H}$}
\begin{document}
   \title{A deep INTEGRAL hard X-ray survey of the 3C~273/Coma region}

   \author{
     S. Paltani \inst{1}
\and
     R. Walter  \inst{1}
\and
     I. M. McHardy  \inst{2}
\and
     T. Dwelly  \inst{2}
\and
     C. Steiner  \inst{1}
\and
     T. J.-L. Courvoisier  \inst{1}
}


   \institute{
ISDC, Geneva Observatory, University of Geneva, ch. d'\'Ecogia 16, CH-1290 Versoix, Switzerland\\
\email{Stephane.Paltani@obs.unige.ch}
\and
School of Physics and Astronomy, University of Southampton, Southampton S017 1BJ, UK
}

   \date{Received 25 January 2008 / Accepted 29 April 2008}

   \abstract{
   We present an analysis of the deepest hard X-ray survey to date of about 2500\,deg$^{2}$ performed by the IBIS instrument on board INTEGRAL in the 20--60\,keV band, with a total exposure time of 4\,Ms. We find 34 candidate sources, for which we try to find counterparts at other wavelengths. The ratio of Seyfert 1 to Seyfert 2 is significantly more than the ratio found in the optical. This effect may be explained in the framework of the receding-torus model, but could also be due to absorption columns large enough to affect the 20--60\,keV band. None of the predicted Compton-thick objects with $10^{24}<$\nh$<10^{25}$\,cm$^{-2}$ is detected unambiguously; when taking lower limits on \nh\ into account, the fraction of these objects is found to be lower than 24\%. We do not see, but cannot exclude, a relationship between absorption and luminosity similar to what is seen in the 2--10\,keV band. Our data suggests the possibility of a lack of objects with $10^{21}\le$\nh$\le 10^{22}$\,cm$^{-2}$, which could be expected if absorption has two origins, for instance a torus-like structure and the host galaxy. We find that the Log\,$N$--Log\,$S$ diagram of our sources is compatible with those obtained in other surveys in hard X-rays. Compared to models of the AGN population selected in the 2--10\,keV band, the Log\,$N$--Log\,$S$ diagram is generally in good agreement, but the \nh\ distribution is significantly different, with significantly less unabsorbed sources (\nh$<10^{22}$\,cm$^{-2}$) at a given flux limit compared to the models. In this survey, we resolve about 2.5\% of the cosmic X-ray background in the 20--60\,keV band. We also study the local hard X-ray luminosity function, which is compatible with what is found in other recent hard X-ray surveys. The characteristic luminosity Log $L^*_{20-60\,\mathrm{keV}}=43.66$ is found to be a factor about 5 lower than the value observed in the 2--10\,keV band. We find a space density of $10^{-3}$ AGN with $L_{20-60\,\mathrm{keV}}>10^{41}$ per Mpc$^3$ and a corresponding luminosity density of $0.9\,10^{39}$\,erg s$^{-1}$ Mpc$^{-3}$.

}



\maketitle
%

\section{Introduction}

Recent progress in the understanding of the cosmological significance of supermassive black holes stirred by both observations \citep{MagoEtal-1998-DemMas,FerrMerr-2000-FunRel} and numerical simulations \citep{KaufHaeh-2000-UniMod,HopkEtal-2005-PhyMod} prompts the making of a census of these objects as complete as possible. At distances too large to allow kinematics studies of stars in central regions of the galaxies, supermassive black holes are best revealed through their accretion of surrounding material. While these active galactic nuclei (AGN) are usually bright over a large fraction of the electromagnetic spectrum, the X-ray domain provides a privileged access to their population, if only for the reason that, at high galactic latitudes, the vast majority of X-ray sources are AGN.

With the advent of powerful X-ray satellites, the 2-10 keV energy range is now very easily accessible. Both deep and wide surveys have been conducted and put together in order to study in detail the X-ray luminosity function of AGN up to cosmological redshifts $z\sim 3$ \citep{UedaEtal-2003-CosEvo,LafrEtal-2005-HarXra} and  higher \citep{SilvEtal-2007-LumFun}, although sample sizes remain very small above $z\sim 3$.

In spite of survey sensitivities in the $\mu$Crab range, it is expected that surveys conducted by XMM-Newton or Chandra may provide a significantly biased view of the AGN population against the most absorbed AGN. \citet{MarsEtal-1980-DifXra} discovered with HEAO-1 an apparently diffuse X-ray emission at high galactic latitude. This so-called cosmic X-ray background presents a prominent peak around 30\,keV, which has long been explained by the presence of highly absorbed AGN, with hydrogen column densities \nh\ larger than $10^{22}$\,cm$^{-2}$, and even Compton-thick objects with \nh$\gtrsim 10^{24}$\,cm$^{-2}$  \citep{SettWolt-1989-ActGal,MadaEtal-1994-UniSey,MattFabi-1994-SpeCon,ComaEtal-1995-ConAGN}. Even when they are intrinsically bright, these Compton-thick objects emit very little radiation below 10\,keV and thus require deep X-ray observations. In a recent detailed modelling of the AGN population based on the known AGN population up to $z\sim 3$ and its extrapolation to higher redshifts, \citet{GillEtal-2007-SynCos} found that the population of Compton-thick AGN should be as large as that of moderately absorbed AGN. In fact a significant fraction of local Seyfert 2 galaxies are found to be likely Compton-thick \citep{RisaEtal-1999-DisAbs,GuaiEtal-2005-XraObs}.

Absorption is much less efficient in the hard X-ray domain ($\gtrsim 20$\,keV) than in lower X-ray bands. Surveys in the hard X-ray have therefore the potential of detecting bright AGN with minimal bias in the \nh\ distribution below $\sim 10^{25}$\,cm$^{-2}$. INTEGRAL and SWIFT are two satellites with such survey capabilities. Thanks to their large fields-of-view, the AGN population could be studied over the full sky with both INTEGRAL \citep{BeckEtal-2006-HarXra,SazoEtal-2007-HarXra} and SWIFT \citep{MarkEtal-2005-SwiHig,TuelEtal-2007-SwiBat}. Unfortunately, these surveys reach very limited sensitivities compared to their lower-energy counterparts. It is therefore very important to keep accumulating exposure time at high latitudes. In the case of INTEGRAL, which has a smaller field-of-view and a science program mostly geared towards the study of galactic sources, this means focusing on a specific, relatively small high-latitude region. Provided a sustained observational effort is deployed over the next years of INTEGRAL operations these observations have however the potential of providing the deepest extragalactic hard X-ray survey for years to come, and is therefore an important goal.

In this paper we study a $\sim 2500$\,deg$^2$ region of the sky centered around 3C~273 and the Coma cluster with the goal of pushing down as much as possible the sensitivity limit of hard X-ray surveys. This region is indeed the high-latitude region of the sky that has been the target of the deepest exposure with INTEGRAL. We study the population properties of the detected AGN, with a focus on their absorption properties. We compare the hard X-ray-selected AGN population with that resulting from medium (2-10\,keV) X-ray surveys. We also investigate the local AGN luminosity function in the hard X-rays.


\section{INTEGRAL observations of the \cc/Coma cluster region}
\subsection{Data and processing}
Several INTEGRAL core-programme and open-time observations have covered the sky region around Coma cluster and 3C~273. We selected all available INTEGRAL pointings within 30 degrees of a position located between these two sources, which resulted in 1660 pointings for a total elapsed observing time of 3\,936\,234\,s and a dead-time corrected good exposure of 2\,733\,202\,s. Most pointings belong to four 5x5 dithering patterns repeated several times, plus a specific rectangular pattern used during the core-programme observation.

Sky images in the 20-60\,keV energy ranges were created from the data taken by the ISGRI detector of the IBIS imager on board INTEGRAL \citep{UberEtal-2003-IbiIma}. Analysis has been performed with the Off-line Science Analysis software \citep[OSA;][]{CourEtal-2003-IntSci}\footnote{http://isdc.unige.ch?Support+documents}, version 6.0, using standard parameters. Good time intervals were built using a strong constraint on the attitude stability (deviation $<2$ arcsec). Image cleaning was based on an input catalog of 34 sources with fixed source positions and was applied independently of the source strength, allowing for negative source model to avoid introducing any bias in the process.

To build the best possible mosaic of the field we excluded the outskirts of individual images, which are noisy and do not add much signal, as well as 10 individual images which had background fluctuations larger than 1.1 in the significance map. 1650 images were finally included in the mosaic, spanning revolutions from 0036 to 0464.

The 3000x3000-pixel mosaic image was built in equatorial coordinates with a tangential projection using a factor-2 over-sampling when compared to the individual input sky images; this results in a pixel size of 2.4 arcmin in the center of the mosaic and of about 1.6 arcmin in the outskirts of the image, roughly 40 degrees away from the center. The photometric integrity and accurate astrometry are obtained by calculating the intersection between input and output pixels and weighting count rates according to the overlapping area.

\subsection{Properties of the mosaic}
\begin{figure}
\fbox{\includegraphics[width=8.8cm]{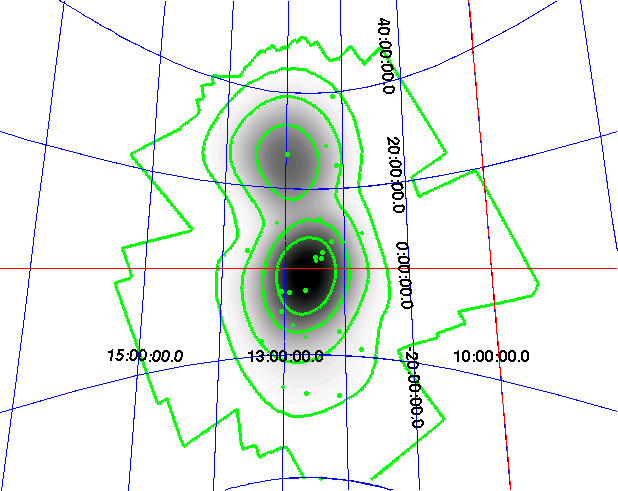}}%
\caption{\label{fig:mosa}
Exposure map of the mosaic around \cc. The contours are located at 0, 10, 100, 300 and 500\,ks respectively. Coordinates are right ascension and declination. Dots indicate the location of detected sources.}
\end{figure}

Fig.~\ref{fig:mosa} shows the exposure map of the field around the 3C~273/Coma region.
The total surface area is about 4900\,deg$^2$. The sky area having been exposed more than $10$\,ks is 2390\,deg$^2$, while 1415\,deg$^2$ have been observed more than $100$\,ks. We point out that all quoted exposure times are effective exposures times, i.e. corrected for loss of efficiency due to dead time and off-axis observations.

\begin{figure}
\includegraphics[width=8.8cm]{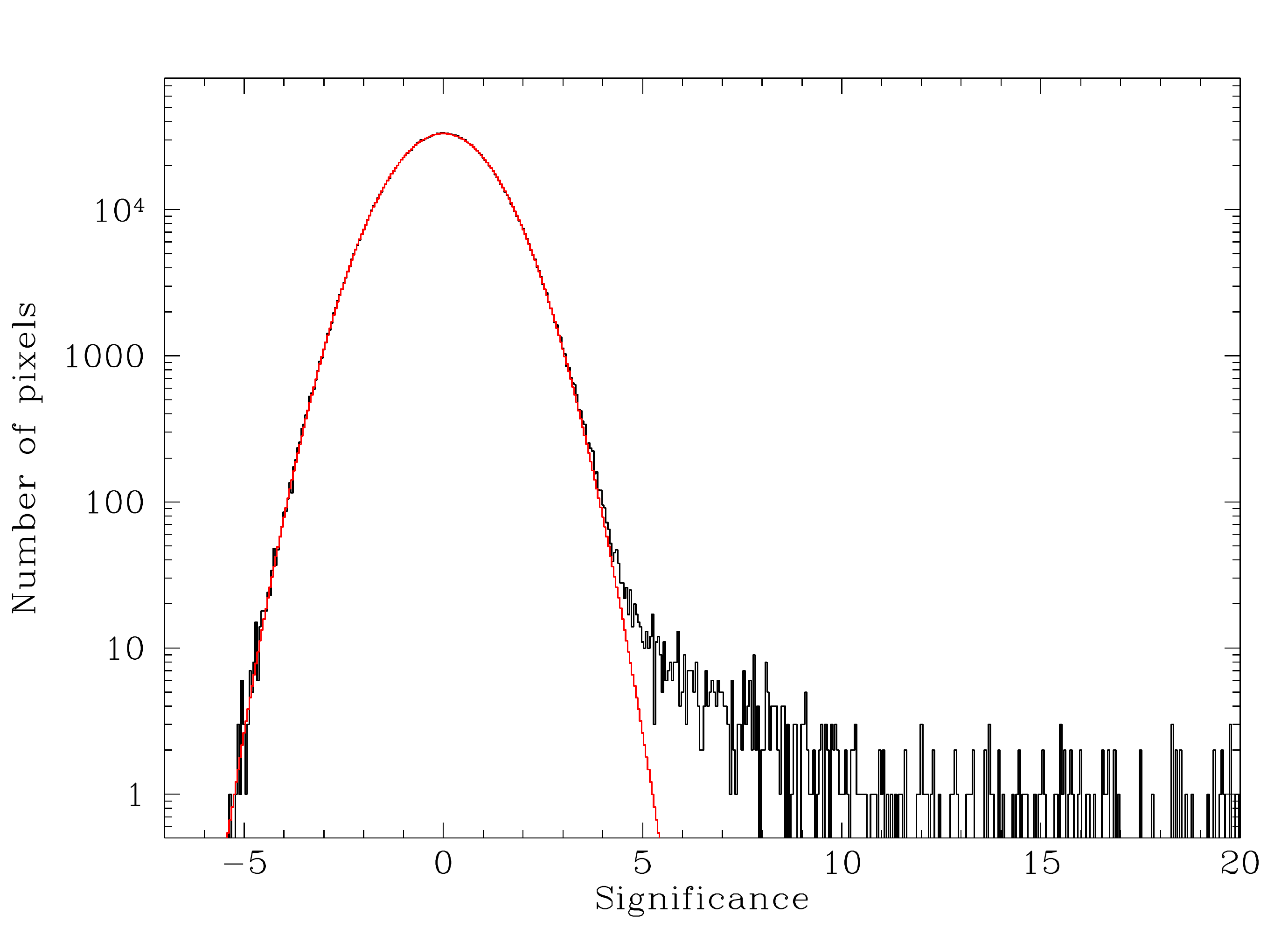}
\caption{\label{fig:sig}
Histogram of the pixels' significance for the part of the mosaic exposed for more than 10\,ks. The red curve is a Gaussian fit to the part of the histogram with $\sigma\le 3$.}
\end{figure}

To investigate the quality of the mosaic, we study the distribution of the pixels' significance, which is expected to be Gaussian with mean 0 and dispersion 1. Significant non-Gaussian tails are present when the full mosaic is studied. Such deviations are expected for sky areas having been observed a small number of times, where systematic effects have not been averaged out. This problem can be avoided by restricting the mosaic to the part where the effective exposure time is longer than 10\,ks. Fig.~\ref{fig:sig} shows the significance distribution of the well-exposed pixels. We fitted a Gaussian distribution to the part of the histogram with $\sigma\le 3$ to avoid the strong positive tail due to real sources. The negative branch of the significance histogram is very well fitted, with no evidence of excess in the tail. The sigma of the Gaussian is about 1.1, significantly larger than the expected value of 1. The centroid of the Gaussian distribution is located at 0.009, which, while significant, has negligible impact on the significance levels of candidate sources.

In view of the deviations found at low exposure times we shall completely discard the part of the mosaic with exposure shorter than 10\,ks. The resulting mosaic contains almost 2 million pixels.

\section{Source extraction and selection}
\subsection{Extraction method}
We build the candidate-source list from the mosaic by selecting all pixels in the mosaic with a significance $\sigma_{\mathrm{pixel}}$ larger than a given threshold small enough to ensure that every significant source has at least one pixel with a larger significance. We used $\sigma_{\mathrm{pixel}}>3$. We use these pixels as starting points for the standard OSA flux-extraction tool, \texttt{mosaic\_spec}. This tool fits a Gaussian peak in a user-defined box centered around the input pixel. We used here a small 12x12-pixel box to avoid that the fitted peak drifts towards nearby, more significant peaks; the box size corresponds to 2.4 times the point-spread function's (PSF) FWHM in the center of the image (where pixels are the smallest). The Gaussian fit is performed letting the centroid free, but fixing the FWHM to that of PSF of ISGRI, i.e. 12 arcmin. As a result, we obtain for each starting-point pixel the coordinates of the peak with its associated uncertainty, a flux (again with its uncertainty), an exposure time and a revised significance $\sigma$, determined by \texttt{mosaic\_spec} on the basis of the variance map. As the fits may converge several times on the same source, we used a distance-based algorithm to remove multiple detections of the same peak.

\subsection{Detection threshold}
\begin{figure}
\includegraphics[width=8.8cm]{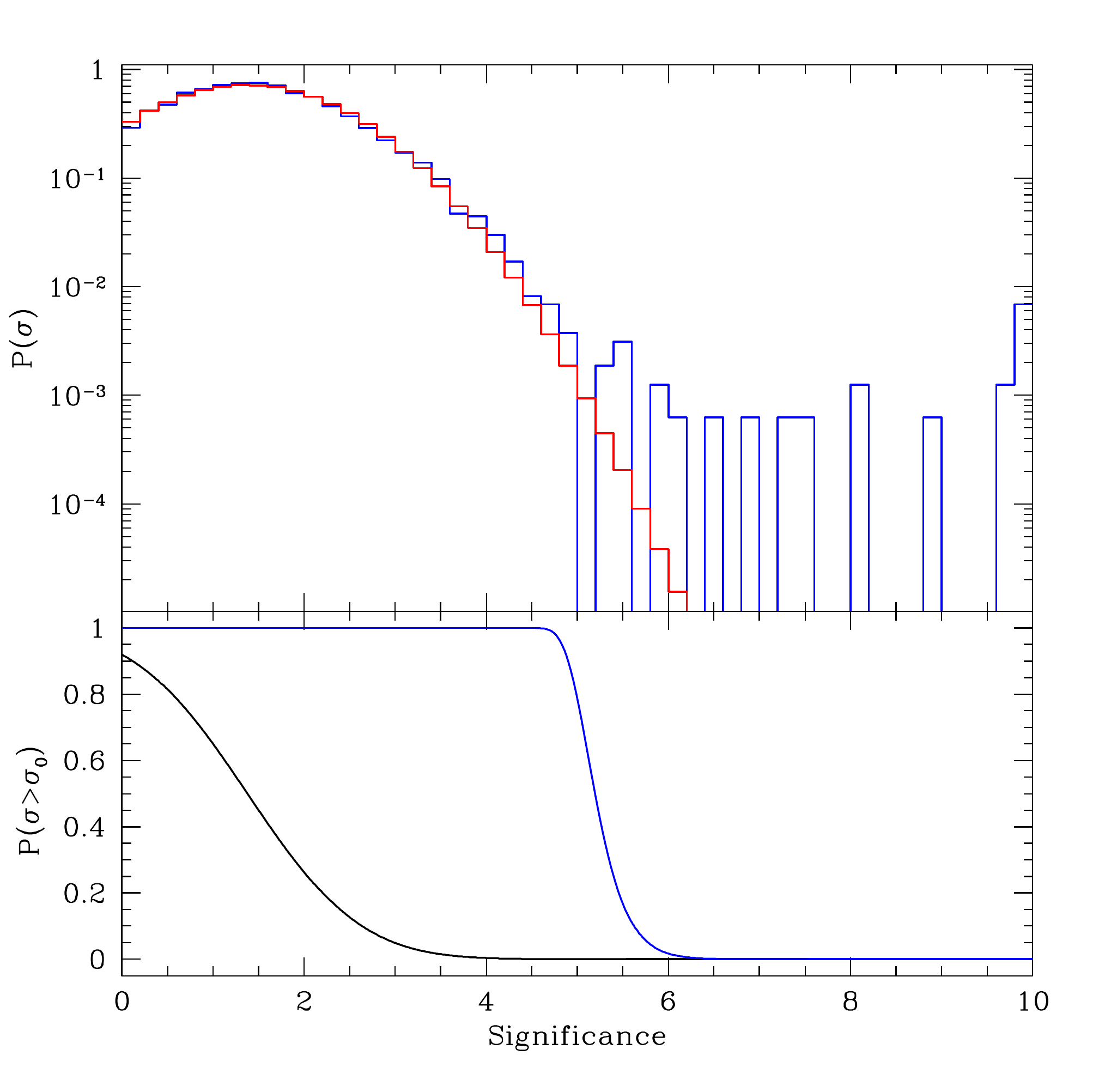}
\caption{\label{fig:sigmhist}
Top: Significance distribution of the pseudo-sources extracted randomly from the mosaic (blue line). The red line is a Gaussian fit to the part with $\sigma<4$. The excesses in the significance distribution at $\sigma>4$ indicate true sources. Bottom: Probability that the significance $\sigma$ of a fake source extracted at a specific position is by chance larger than a given significance $\sigma_0$ (black line). The blue line gives the probability to find a fake source with significance $\sigma>\sigma_0$ anywhere in the mosaic.}
\end{figure}

\begin{figure}
\includegraphics[width=8.8cm]{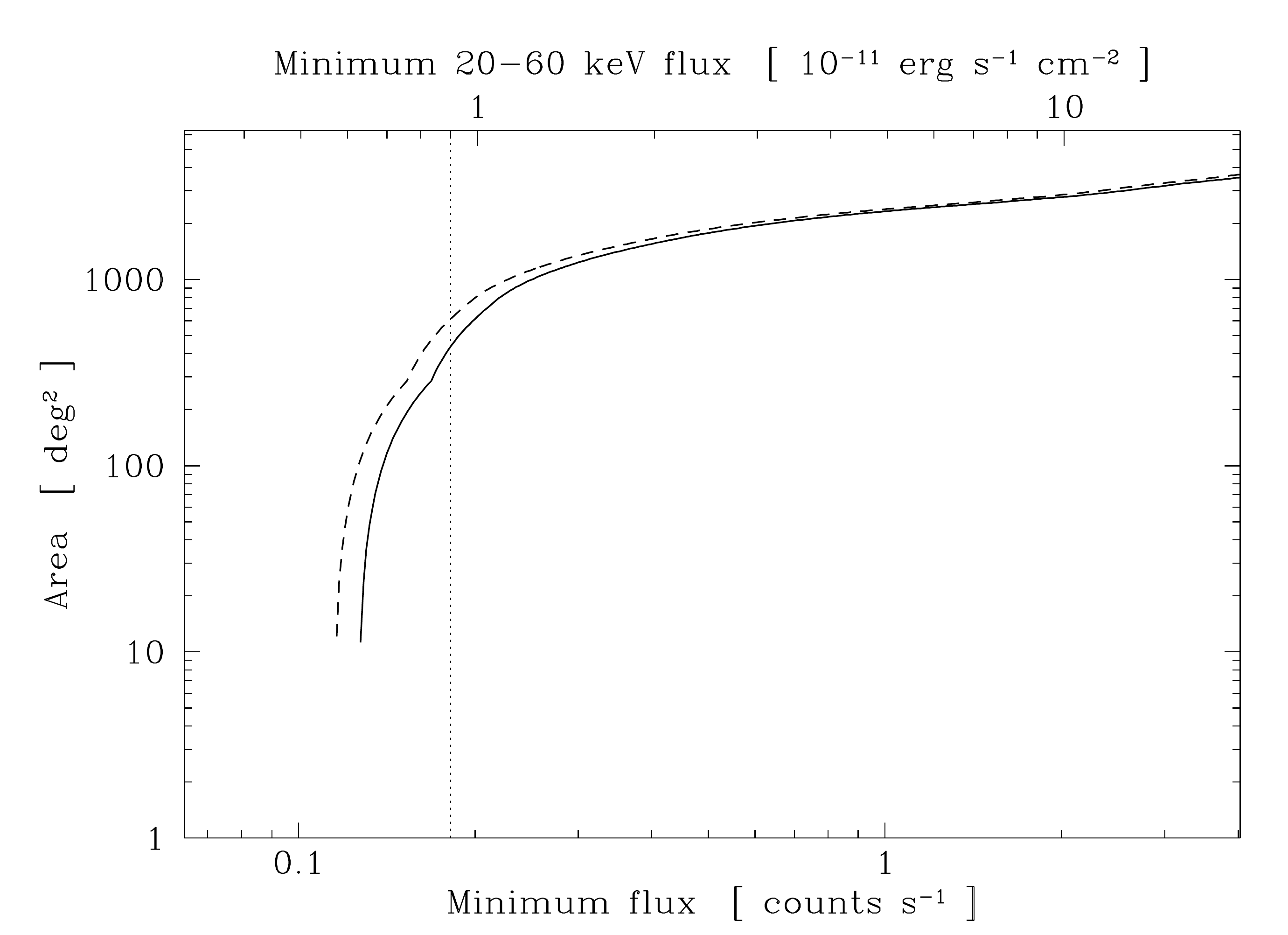}
\caption{\label{fig:fmin}
Surface of the sky over which a given flux results in a significance $\sigma\ge 5.5$ (solid line) and $\sigma\ge 5.0$ (dashed line). The vertical dotted line shows the sensitivity limit of other hard X-ray surveys.}
\end{figure}

Detection threshold obviously depends on the input mosaic, but it also depends on the exact method used to select and extract sources. In order to determine the detection threshold for the candidate sources, we investigate the distribution of the significance $\sigma$ obtained if we apply the above extraction method starting from a random point on the mosaic. Fig.~\ref{fig:sigmhist} shows the significance distribution of the pseudo-sources obtained by drawing a large number of starting points at random over the part of the mosaic with more than 10\,ks exposure. After removal of duplicates, we estimate that there are about 13\,800 unique possible sources over the full mosaic. Their significance distribution follows very well a Gaussian distribution, with a strong positive tail due to the fit occasionally converging on real sources. We fit the part of the distribution with significance $\sigma<4$ with a Gaussian. The centroid of the Gaussian model is located at $1.38$, and the variance is compatible with $1$. Thus, using the Gaussian approximation, one can derive the probability that source extraction at a fixed starting position returns a significance larger than a given significance $\sigma_0$ in absence of a source at this position. However, when looking for a priori unknown sources, we have to take into account that there are 13\,800 unique sources in the mosaic. The probability to find a source with a significance $\sigma>\sigma_0$ is therefore given by the binomial distribution of the fixed-source probability with 13\,800 repetitions. The two probability cumulative distributions are shown in Fig.~\ref{fig:sigmhist}. The probability that a new candidate source with a significance $\sigma=5$ is real is about 20\%. It increases to 85\% for $\sigma=5.5$ and to more than 98\% for $\sigma=6$. We point out that a number of candidate sources with significance just below the $\sigma=5$ limit might turn out to be real sources.
Using this probability distribution, we can now calculate the sky area over which a source with a given flux can be detected. An example is given in Fig.~\ref{fig:fmin} for a significance threshold $\sigma=5.5$. At the significance level $\sigma=5.5$, the survey reaches a depth of 1 count s$^{-1}$ over the totality of the 2390\,deg$^2$ with more than 10\,ks exposure.

\subsection{Properties of the hard X-ray sources}
\label{sec:src}
\subsubsection{Source identification}
Tables~\ref{tab:sources} and \ref{tab:prop} list the 34 candidate sources detected in the mosaic with a significance $\sigma>5$ together with their identifications and basic properties, when available. Source identification is performed by searching for a counterpart in a radius depending on the uncertainty on the source position. When the source cannot be obviously related to a known counterpart, we perform a tentative identification using both Simbad database\footnote{http://simbad.u-strasbg.fr/simbad} from CDS (Centre de Donn\'ees de Strasbourg) and the NASA Extragalactic Database\footnote{NED; http://nedwww.ipac.caltech.edu}. Counterparts in the error circles are selected based on the following criteria, in order of priority: (1) low-redshift AGN; (2) X-ray sources; (3) low-redshift galaxies; (4) Infrared or radio sources. Once a counterpart is chosen, class and redshift information is taken from Simbad or NED, unless specified in Table~\ref{tab:prop}.

To quantify the probability of finding counterparts of a given type, we drew 100 random coordinates over the mosaic and searched for counterparts in a 3.5 arcmin radius using Simbad; this radius corresponds to the typical 90\% position uncertainty of the least significant candidate sources. Table~\ref{tab:prob_coinc} gives the probabilities we found for different category sources. The probability of chance coincidence with an AGN is very low, especially if we restrict the redshift range to $z<0.1$, which means that probably all candidate sources identified with low-redshift AGN are true detections. We found however non-negligible probabilities of coincidence with apparently normal low-redshift galaxies, X-ray, infrared and radio counterparts, showing that such matches do not guarantee that the detection is not fake. In practice, we shall only consider that matches with low-redshift AGN indicate a secure detection.
\begin{table}[tb]
\caption{Probability of chance coincidence for the candidate sources with counterparts in other catalogs.}
\begin{center}
\begin{tabular}{lr}
\hline
\rule{0pt}{1.2em}%
Object type&Probability\\
\hline
\rule{0pt}{1.2em}%
AGN ($z<0.1$)& 1\,\%\\
AGN (all) & 3\,\%\\
X-ray sources & 5\,\%\\
Infrared sources & 12\,\%\\
Radio sources & 13\,\%\\
Galaxies & 26\,\%\\
\hline
\end{tabular}
\end{center}
\label{tab:prob_coinc}
\end{table}%

\begin{table*}[t]
\caption{Properties of the candidate sources detected in the mosaic. Coordinates are those of the excesses in the mosaic. $\varepsilon$ is the 90\% uncertainty on the position and $\delta$ is the distance between the sources. Identification and classes are taken from Simbad or NED, unless specified in Table~\ref{tab:prop}. $\sigma$ is the source's significance. $P$ is the probability that the source is real, not taking into account the identification. Star indicate a match with with low-redshift AGN or otherwise unquestionable AGN counterparts.}
\begin{center}
\begin{tabular}{lccccllccc}
\hline
\multicolumn{1}{c}{Name}&\rule{0pt}{1.2em}RA&Dec& $\varepsilon$& $\delta$ &\multicolumn{1}{c}{Identification}&\multicolumn{1}{c}{Class}&Exposure&$\sigma$&$P$\\
&\multicolumn{2}{c}{J2000.0}&\multicolumn{2}{c}{arcmin}&&&s\\
\hline \rule{0pt}{1.2em}%
IGR J12291+0203&12 29 07&+\z 2 03 03&0.20&0.19& \object{3C 273} & Blazar &699\,100&97.69&\phantom{*}1.00*\\
IGR J12258+1240&12 25 46&+12 39 45&0.35&0.19& \object{NGC 4388} & Seyfert 2 &278\,664&79.90&\phantom{*}1.00*\\
IGR J12106+3925&12 10 34&+39 24 39&0.43&0.38& \object{NGC 4151} & Seyfert 1.5 &\z 21\,845&42.77&\phantom{*}1.00*\\
IGR J12396$-$0521&12 39 37&$-$\z 5 21 02&1.64&0.39& \object{NGC 4593} & Seyfert 1 &726\,431&38.03&\phantom{*}1.00*\\
IGR J12562$-$0547&12 56 12&$-$\z 5 46 33&1.70&0.87& \object{3C 279} & Blazar &629\,183&10.98&\phantom{*}1.00*\\
IGR J12392$-$1612&12 39 10&$-$16 11 41&1.61&1.29& \object{LEDA 170194} & Seyfert 2 &249\,172&10.64&\phantom{*}1.00*\\
IGR J12595+2755&12 59 28&+27 54 53&2.07&6.04& \object{Coma cluster} & Gal. cluster &429\,197&10.42&1.00\\
IGR J12233+0241&12 23 21&+\z 2 41 05&1.82&0.92& \object{MRK 50} & Seyfert 1 &636\,864&\z 9.69&\phantom{*}1.00*\\
IGR J12226+0414&12 22 36&+\z 4 13 34&1.96&3.31& \object{4C 04.42} & Blazar &590\,488&\z 9.68&\phantom{*}1.00*\\
IGR J13092+1137&13 09 10&+11 37 26&1.78&1.31& \object{NGC 4992} & XBONG &185\,410&\z 9.66&\phantom{*}1.00*\\
IGR J12390$-$2719&12 38 57&$-$27 18 57&2.06&0.81& \object{MCG$-$04$-$30$-$007} & Seyfert 2 &\z 34\,443&\z 8.98&\phantom{*}1.00*\\
IGR J13383+0434&13 38 17&+\z 4 33 51&2.59&1.32& \object{NGC 5252} & Seyfert 1.5 &\z 58\,992&\z 7.54&\phantom{*}1.00*\\
IGR J12185+2948&12 18 28&+29 48 27&2.52&0.44& \object{NGC 4253} & Seyfert 1.5 &211\,282&\z 6.71&\phantom{*}1.00*\\
IGR J13225$-$1645&13 22 29&$-$16 44 41&2.95&1.51& \object{MCG$-$03$-$34$-$064} & Seyfert 1.8 &112\,932&\z 6.55&\phantom{*}1.00*\\
IGR J13041$-$0533&13 04 05&$-$\z 5 32 52&2.77&2.04& \object{NGC 4941} & Seyfert 2 &534\,821&\z 6.13&\phantom{*}0.99*\\
IGR J12299+0305&12 29 53&+\z 3 04 57&3.67&5.58& \object{1RXS J123013.6+030258} & X-ray source &677\,378&\z 5.92&0.97\\
IGR J12069$-$1448&12 06 54&$-$14 47 42&3.59&1.18& \object{2MASX J12065497$-$1446335} & Galaxy &147\,802&\z 5.60&0.89\\
IGR J13042$-$1020&13 04 13&$-$10 19 58&3.14&0.60& \object{NGC 4939} & Seyfert 2 &414\,987&\z 5.52&\phantom{*}0.84*\\
IGR J12174$-$0131&12 17 27&$-$\z 1 31 19&4.20&2.64& \object{FIRST J121735.9$-$013001} & Radio source &616\,797&\z 5.49&0.83\\
IGR J12011+0649&12 01 03&+\z 6 48 43&3.23&1.37& \object{LEDA 37894} & Seyfert 2 &286\,255&\z 5.49&\phantom{*}0.82*\\
IGR J13415+3023&13 41 32&+30 23 24&3.50&4.49& \object{MRK 268} & Seyfert 2 &189\,232&\z 5.43&\phantom{*}0.78*\\
IGR J12042$-$0756&12 04 11&$-$\z 7 55 44&3.79&1.62& \object{LEDA 157316} & Galaxy &311\,550&\z 5.38&0.73\\
IGR J12070+2535&12 07 03&+25 34 57&3.42&3.75& \object{IRAS 12046+2554} & Galaxy &141\,198&\z 5.37&0.72\\
IGR J13133$-$1109&13 13 17&$-$11 08 34&3.59&2.84& \object{1RXS J131305.9$-$110731} & Seyfert 1 &308\,838&\z 5.36&\phantom{*}0.71*\\
IGR J12060+3818&12 06 02&+38 17 48&3.96&2.80& \object{2MASX J12055104+3819308} & Infrared &\z 23\,423&\z 5.26&0.59\\
IGR J13517$-$0042&13 51 44&$-$\z 0 42 27&4.09&3.40& \object{SDSS J135133.23$-$004024.3} & Galaxy &\z 40\,371&\z 5.21&0.52\\
IGR J12130+0701&12 13 01&+\z 7 01 22&2.86&1.03& \object{NGC 4180} & Seyfert 2 &420\,061&\z 5.20&\phantom{*}0.50*\\
IGR J11457$-$1827&11 45 41&$-$18 27 29&3.44&0.26& \object{1H 1142$-$178} & Seyfert 1 &\z 30\,405&\z 5.10&\phantom{*}0.36*\\
IGR J12172+0710&12 17 09&+\z 7 09 33&3.31&1.92& \object{NGC 4235} & Seyfert 1 &455\,089&\z 5.09&\phantom{*}0.35*\\
IGR J12136$-$0527&12 13 37&$-$\z 5 26 53&3.84&5.15& \object{1RXS J121353.4$-$053000} & X-ray source &509\,473&\z 5.08&0.33\\
IGR J12310+1221&12 31 00&+12 21 28&3.57&3.26& \object{M 87} & Radiogalaxy &299\,332&\z 5.08&\phantom{*}0.33*\\
IGR J11225$-$0419&11 22 31&$-$\z 4 19 17&5.07&3.50& \object{2MASX J11222455$-$0416096} & Galaxy &\z 21\,144&\z 5.05&0.28\\
IGR J13353$-$1113&13 35 17&$-$11 13 10&3.96&6.00& \object{NVSS J133453$-$111300} & Radio source &106\,440&\z 5.04&0.27\\
IGR J11427+0854&11 42 40&+\z 8 54 29&3.92&2.10& \object{2MASX J11424200+0852251} & Galaxy &\z 68\,842&\z 5.03&0.25\\
\hline
\end{tabular}
\end{center}
\label{tab:sources}
\end{table*}%
\begin{table*}[t]
\caption{Properties of the candidate sources detected in the mosaic. Fluxes and luminosities are in the 20-60\,keV energy range. $m_{\mathrm B}$ is the apparent magnitude in filter $B$, unless specified in the footnote. \nh\ is the intrinsic hydrogen column density; values in parenthesis are derived from the RASS-BSC fluxes; lower limits mean non-detection in the ROSAT-BSC catalog. Unless mentioned in the comments, redshifts and class information have been obtained through Simbad or NED.  Star indicate a match with with low-redshift AGN or otherwise unquestionable AGN counterparts.}
\begin{center}
\begin{tabular}{lcccccl}
\hline
\multicolumn{1}{c}{\rule{0pt}{1.2em}Name}&Flux&$m_{\mathrm B}$&\nh&Redshift&$\log_{10}L_{\mathrm{obs}}$&Comments\\
&$10^{-11}$ erg\,s$^{-1}$\,cm$^{-2}$&&$10^{22}$ cm$^{-2}$&&erg s$^{-1}$\\
\hline \rule{0pt}{1.2em}%
IGR J12291+0203*& 11.56 $\pm$ 0.12&13.1& $<\!\!\phantom{0}0.01\phantom{>}$ & 0.1583 & 45.97 & {\scriptsize \nh: \citet{ShinEtal-2006-SpeSta}}\\
IGR J12258+1240*& 15.27 $\pm$ 0.19&12.2& 26.92 & 0.0084 & 43.38 & {\scriptsize \nh: \citet{ShuEtal-2007-InvNuc}}\\
IGR J12106+3925*& 30.74 $\pm$ 0.72&11.2& 7.5 & 0.0033 & 42.87 & {\scriptsize \nh: \citet{CappEtal-2006-XraSpe}}\\
IGR J12396$-$0521*&\z 4.48 $\pm$ 0.12&13.5& $<\!\!\phantom{0}0.01\phantom{>}$ & 0.0090 & 42.91 & {\scriptsize \nh: \citet{ShinEtal-2006-SpeSta}}\\
IGR J12562$-$0547*&\z 1.37 $\pm$ 0.12&18.0& \z 0.05 & 0.5362 & 46.39 & {\scriptsize \nh: \citet{ReevTurn-2000-XraSpe}}\\
IGR J12392$-$1612*&\z 2.24 $\pm$ 0.21&10.9& 1.9 & 0.036\z & 43.84 & {\scriptsize Class: \citet{MaseEtal-2006-UnvNatII}; \nh: \citet{SazoEtal-2005-IdeEig}}\\
IGR J12595+2755&\z 1.52 $\pm$ 0.15&13.5& $<\!\!\phantom{0}0.01\phantom{>}$ & 0.0231 & 43.28 & {\scriptsize \nh: \citet{ArnaEtal-2001-XmmObs}}\\
IGR J12233+0241*&\z 1.20 $\pm$ 0.12&15.5& (\phantom{0}0.25) & 0.0234 & 43.18 \\
IGR J12226+0414*&\z 1.23 $\pm$ 0.13&17.5& (\phantom{0}7.1\phantom{0}) & 0.9650 & 47.08 \\
IGR J13092+1137*&\z 2.36 $\pm$ 0.24&14.6& 90.0\z & 0.0251 & 43.54 & {\scriptsize Class: \citet{MaseEtal-2006-UnvNatIV}; \nh: \citet{SazoEtal-2005-IdeEig}}\\
IGR J12390$-$2719*&\z 5.48 $\pm$ 0.61&14.6& 10\phantom{.00} & 0.0250 & 43.90 & {\scriptsize Class: \citet{MoreEtal-2006-OptCla}; \nh: \citet{TuelEtal-2005-SwiDet}}\\
IGR J13383+0434*&\z 3.25 $\pm$ 0.43&14.5& \z 2.89 & 0.0230 & 43.60 & {\scriptsize \nh: \citet{ShuEtal-2007-InvNuc}}\\
IGR J12185+2948*&\z 1.45 $\pm$ 0.22&14\phantom{.0}& $<\!\!\phantom{0}0.02\phantom{>}$ & 0.0129 & 42.74 & {\scriptsize \nh: \citet{PageEtal-2001-VarXmm}}\\
IGR J13225$-$1645*&\z 2.02 $\pm$ 0.31&14\phantom{.0}& 73.85 & 0.0165 & 43.10 & {\scriptsize \nh: This work (Chandra); see text}\\
IGR J13041$-$0533*&\z 0.84 $\pm$ 0.14&12\phantom{.0}& 44.98 & 0.0037 & 41.41 & {\scriptsize \nh: \citet{ShinEtal-2006-SpeSta}}\\
IGR J12299+0305&\z 0.71 $\pm$ 0.12&-& (\phantom{0}2.5\phantom{0}) & - & - & {\scriptsize QSO \object{2MASS J12301552+0302546}, $z=0.138$, $m_{\mathrm V}=17.5$ ?}\\
IGR J12069$-$1448&\z 1.55 $\pm$ 0.28&-& $>2$ & 0.0184 & 43.08 \\
IGR J13042$-$1020*&\z 0.87 $\pm$ 0.16&11\phantom{.0}& 30.0\z & 0.0104 & 42.33 & {\scriptsize \nh: \citet{BassEtal-1999-ThrDia}}\\
IGR J12174$-$0131&\z 0.70 $\pm$ 0.13&-& $>2$ & - & - & {\scriptsize No obviously associated optical or infrared source}\\
IGR J12011+0649*&\z 1.04 $\pm$ 0.19&15.3& \z 6.61 & 0.0360 & 43.51 & {\scriptsize Class: \citet{MaseEtal-2006-OptCla}; \nh: \citet{LandEtal-2007-AGNNat}}\\
IGR J13415+3023*&\z 1.25 $\pm$ 0.23&15.3& $>2$ & 0.0399 & 43.68\\
IGR J12042$-$0756&\z 1.00 $\pm$ 0.19&-& $>2$ & - & - & {\scriptsize $m_{\mathrm J}=14.1$, $m_{\mathrm K}=13.0$}\\
IGR J12070+2535&\z 1.42 $\pm$ 0.27&19.9& $>2$ & 0.0477 & 43.90 \\
IGR J13133$-$1109*&\z 0.99 $\pm$ 0.19&15.6& (\phantom{0}0.32) & 0.0344 & 43.45 \\
IGR J12060+3818&\z 3.75 $\pm$ 0.71&-& $>2$ & - & -& {\scriptsize $m_{\mathrm J}=15.5$, $m_{\mathrm K}=14.3$}\\
IGR J13517$-$0042&\z 2.70 $\pm$ 0.52&19.2& $>2$ & 0.0531 & 44.28 \\
IGR J12130+0701*&\z 0.81 $\pm$ 0.16&13.2& $>2$ & 0.0070 & 41.95 & {\scriptsize Class: \citet{MaseEtal-2007-SpeSix}}\\
IGR J11457$-$1827*&\z 3.29 $\pm$ 0.65&14.6& ($<\!\!\phantom{0}0.01\phantom{>}$) & 0.0329 & 43.93 \\
IGR J12172+0710*&\z 0.76 $\pm$ 0.15&13.2& \z 0.15 & 0.0080 & 42.04 & {\scriptsize \nh: \citet{MaseEtal-2006-UnvNatIV}}\\
IGR J12136$-$0527&\z 0.71 $\pm$ 0.14&-& (\phantom{0}2.0\phantom{0}) & - & - & {\scriptsize Sy 1 \object{2MASX J12135456-0530193}, $z=0.066$, $m_{\mathrm V}=16.2$ ?}\\
IGR J12310+1221*&\z 0.94 $\pm$ 0.19&10.4& \z 0.02 & 0.044\z & 43.65 & {\scriptsize \nh: \citet{LieuEtal-1996-DisMil}}\\
IGR J11225$-$0419&\z 3.54 $\pm$ 0.70&17.5& $>2$ & 0.0535 & 44.41 \\
IGR J13353$-$1113&\z 1.59 $\pm$ 0.31&-& $>2$ & - & - & {\scriptsize No obviously associated optical or infrared source}\\
IGR J11427+0854&\z 2.08 $\pm$ 0.41&\phantom{$^1$}15.3$^1$& $>2$ & 0.0213 & 43.34 \\
\hline
\end{tabular}\\
\hspace*{3mm}$^1$ $g$ band\rule{0pt}{1.2em}\hfill ~
\end{center}
\label{tab:prop}
\end{table*}%

\subsubsection{Flux and luminosity}
\label{sec:prop}
We calculate the source flux by fitting a standard AGN spectrum to the 20-60\,keV count rates extracted from the mosaic. We adopted a cut-off power-law with a slope $\Gamma=1.9$ and a cut-off energy $E_{\mathrm{C}}=200$\,keV \citep{PeroEtal-2002-ComRef}. We tested different choices of model parameters, for instance changing $\Gamma$ by $\pm 0.1$ or setting $E_{\mathrm{C}}$ to 100\,keV or above 200\,keV;  the difference in flux has been found to be about 5\%, which can be considered negligible here. We find that a count rate of 1\,s$^{-1}$ corresponds to a flux of $4.95\,10^{-11}$ 
erg\,s$^{-1}$\,cm$^{-2}$, which is about 5\,mCrab. When redshift information is available, luminosity is derived from the luminosity distance, using $H_0=70$\,km\,s$^{-1}$\,Mpc$^{-1}$, $\Omega_{\mathrm M}=0.3$ and $\Omega_{\Lambda }=0.7$.

For comparison with other surveys in the hard X-rays, the 20--60\,keV flux of a source with a spectrum given by the above power-law is 1.55 higher than the 20--40\,keV flux used in \citet{BeckEtal-2006-HarXra}; the 17--60\,keV flux used in \citet{SazoEtal-2007-HarXra} is 15\% higher than the 20--60\,keV flux; and the 14--195\,keV \citep{TuelEtal-2007-SwiBat} is 2.16 times higher.

Using these flux conversion factors, we find that previous hard X-ray surveys \citep{BeckEtal-2006-HarXra,SazoEtal-2007-HarXra,TuelEtal-2007-SwiBat} have all equivalent zero-area sensitivities in the $20-60\mathrm{\,keV}$ domain of about $0.9\,10^{-11}$ erg\,s$^{-1}$\,cm$^{-2}$. At this flux limit, our survey still covers about 450\,deg$^2$, i.e. 15\% of the full mosaic (see Fig.~\ref{fig:fmin}), while the ultimate depth of our mosaic is about $0.6\,10^{-11}$ erg\,s$^{-1}$\,cm$^{-2}$. On the other hand, these surveys cover a much larger fraction of the sky ($>75$\%) than ours ($6$\%).

\subsubsection{Determination of the intrinsic absorption}
\begin{figure}
\includegraphics[width=6.2cm,angle=90]{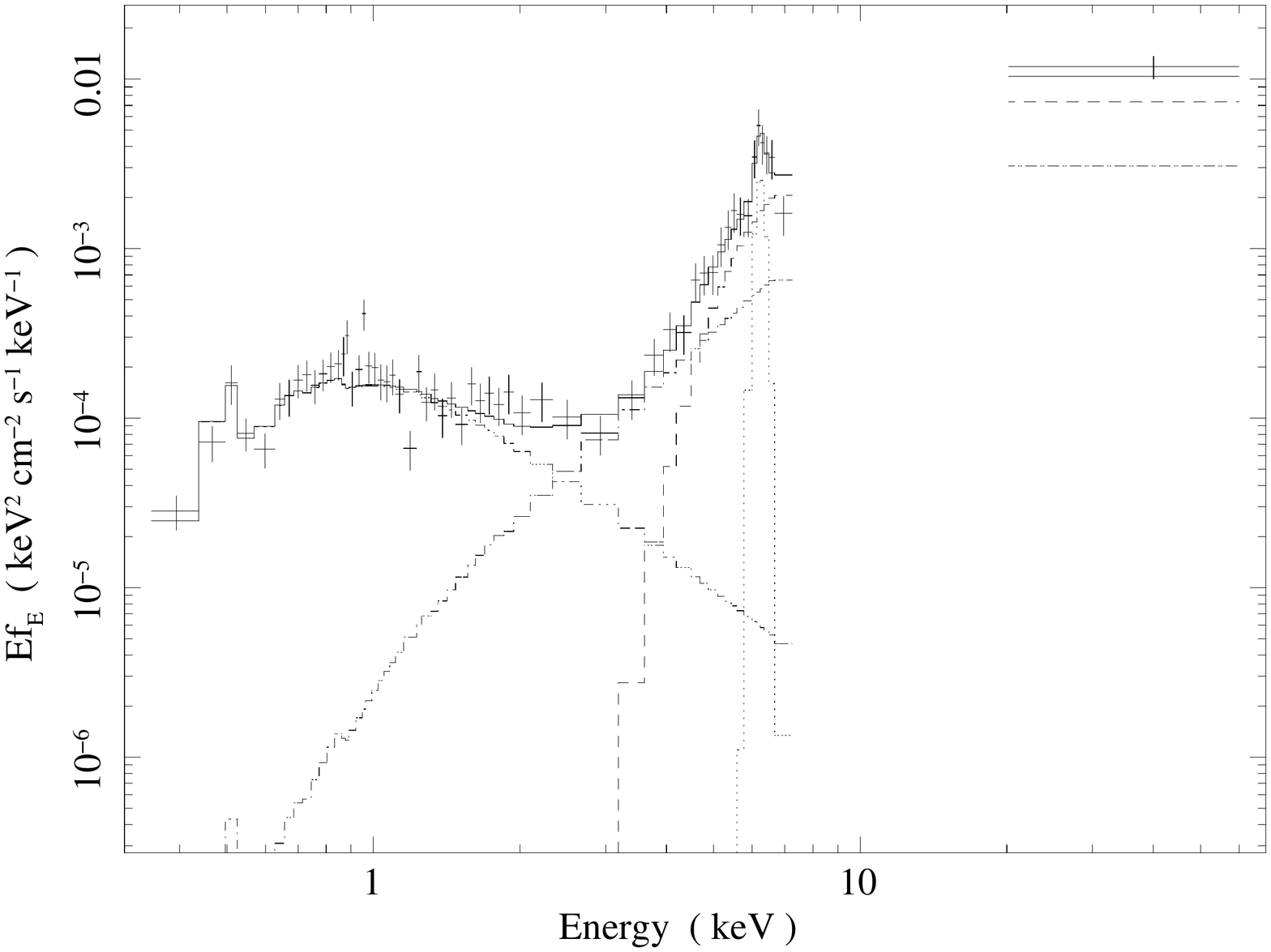}
\caption{\label{fig:mcg}
Combined Chandra and ISGRI spectrum of MCG$-$03$-$34$-$064 in the 0.3--60 keV energy range. The source is well fitted by a model (solid line) consisting of a very absorbed power-law (dashed line) plus a reflection component (dot-dot-dashed line) and an Fe line (dotted line). a very steep component, probably due to scattered X-ray emission, is visible below 2\,keV (dot-dot-dot-dashed line). }
\end{figure}

We searched the literature for previous X-ray measurements of the identified sources in the mosaic, in order to quantify their intrinsic hydrogen column densities \nh. For some objects, no absorption is detected beyond that originating from galactic hydrogen column density. In these cases, we list upper limits at the level of the galactic \nh. We note that, our field being at high galactic latitude, these values are always on the order of $10^{20}$\,cm$^{-2}$. One of these objects, MCG$-$03$-$34$-$064, does not have a published spectrum, although it has been observed serendipitously by Chandra (Observation ID: 7373). We extracted and analyzed the spectrum, which turned out to be strongly absorbed, albeit still in the Compton-thin regime (\nh$\simeq 0.74\,10^{24}$\,cm$^{-2}$). The spectrum is presented in Fig.~\ref{fig:mcg}.

\begin{figure}
\includegraphics[width=8.8cm]{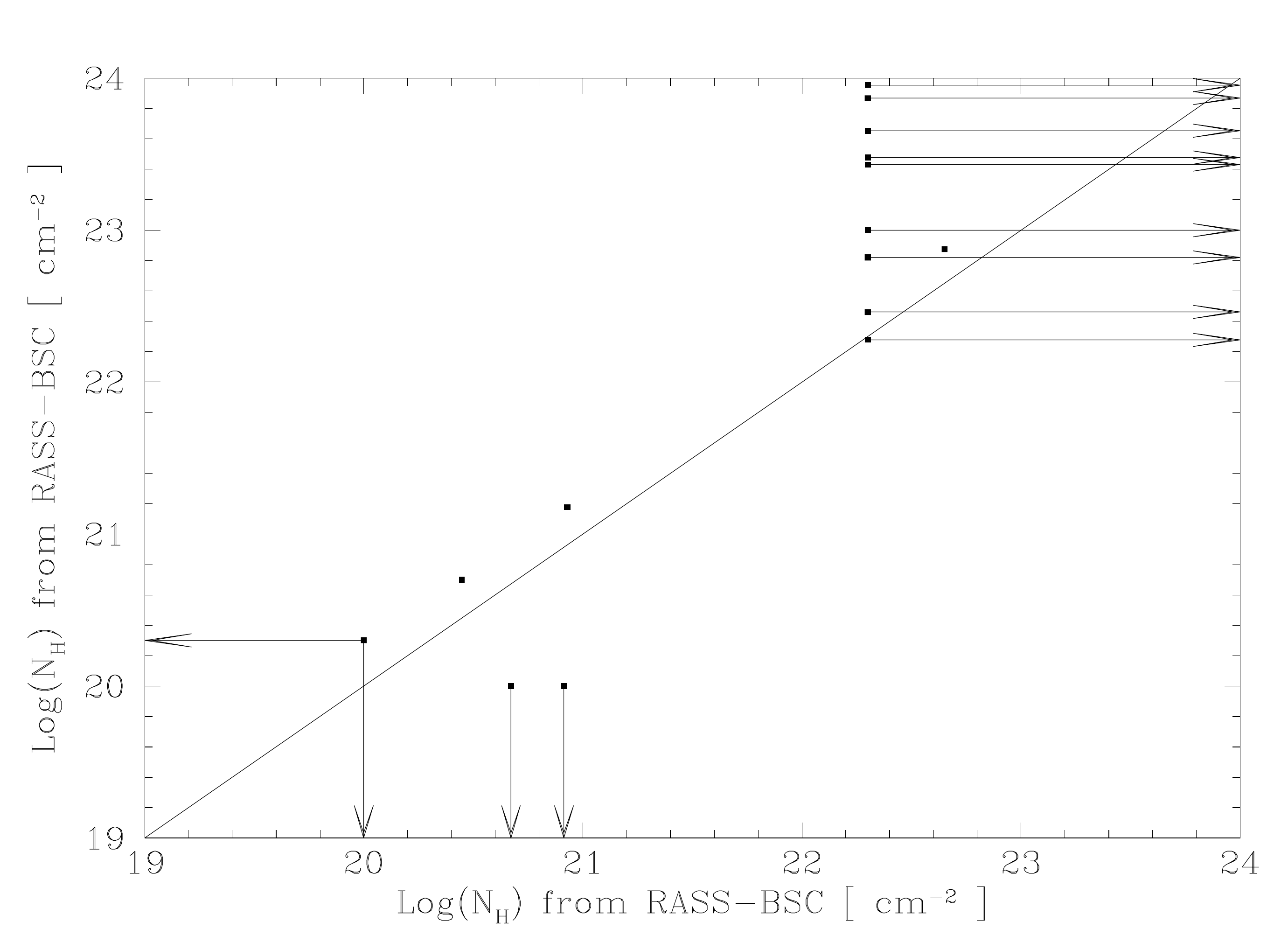}
\caption{\label{fig:rosnh}
Relationship between \nh\ measured in X-ray observations and that derived from the counterpart in RASS-BSC. Coma and M87 are dominated by thermal emission in the soft X-ray band, and are therefore excluded.}
\end{figure}

When no adequate X-ray measurement exists, we check whether there is a counterpart in the ROSAT all-sky survey bright source catalog \citep[RASS-BSC;][]{VogeEtal-1999-ROSBSC}. This catalog is quasi-complete over the full sky down to a flux of 0.05\,cts\,s$^{-1}$ in the 0.1--2.4\, keV band. Using the web version of HEASARC's PIMMS\footnote{\texttt{http://heasarc.gsfc.nasa.gov/Tools/w3pimms.html}} we calculated the expected count rate in the ROSAT band for different values of intrinsic \nh\ assuming a power-law intrinsic emission with index $\Gamma=1.9$, taking into account galactic absorption and source redshift. The method's accuracy is limited by effects of source variability, presence of soft-excess X-ray emission and extrapolation of a power-law whose actual index is unknown, but it is sufficient to obtain a moderately accurate estimate of \nh. Fig.~\ref{fig:rosnh} shows that RASS-BSC-derived \nh\ are estimated with an accuracy better than a factor 2. More importantly, it is able to identify correctly all the absorbed AGN (\nh$>10^{22}$\,cm$^{-2}$).

Absorption in the ROSAT 0.1--2.4\,keV band increases rapidly at hydrogen column densities around $10^{22}$\,cm$^{-2}$. Therefore the presence of a counterpart in RASS-BSC implies \nh$\leq 2\,10^{22}$\,cm$^{-2}$, depending quite moderately on the exact spectral shape. When no counterpart is present in RASS-BSC, we can constrain \nh\ by imposing that the expected count rate in the ROSAT energy band be below 0.05\,cts\,s$^{-1}$, which translates into a lower limit \nh$>2\,10^{22}$\,cm$^{-2}$. For the objects for which we have no redshift, we have to make the further assumption that they lie at low redshift $z\ll 1$, otherwise this lower limit should be raised in their case. The case of 4C\,04.42 shows that, should one of these sources have a redshift $z\sim 1$, the lower limit could be as high as $10^{23}$\,cm$^{-2}$. Caution must however be taken for candidate sources without X-ray counterparts in the RASS-BSC, since fake detections have the highest chance of falling into this category.

\section{The population of hard X-ray selected AGN}

\subsection{The mix of AGN types in hard X-ray selected samples}
In Table~\ref{tab:sources} we list all the sources in our observed area down to a significance level of $\sigma=5$. As discussed earlier, because of the statistical properties of the mosaic, at this significance level there is a fairly large probability that the source is fake. As we are observing a field of high galactic latitude in a hard energy band, AGN are by far the most likely counterparts, with the exception of the Coma cluster \citep{EckeEtal-2007-SouWes}. Conversely, association with a low-redshift AGN indicates that the candidate source is most probably real. From hereon in this section we consider only the 22 sources labeled with a star.

Three blazars are very clearly detected, to which the possible identification of the radiogalaxy M~87 can be added. As \cc\ was the target of most of the observations, it is impossible to draw any statistics on the population of these objects. However, their redshift distribution is very different from that of the rest of the sources, as all blazars have redshifts higher than $0.1$, reaching almost $z\simeq 1$, while the highest redshift for a Seyfert galaxy is $0.04$.

Out of the 18 remaining candidate sources, 17 could be identified as Seyfert galaxies. Five of these Seyfert galaxies are classified as Seyfert 1's and eight as Seyfert 2's. The remaining Seyfert galaxies have intermediate types \citep{Oste-1981-SeyGal}: 3 Seyfert 1.5 and one Seyfert 1.8. The fraction of Seyfert galaxies selected using optical spectroscopy has been determined by \citet{HaoEtal-2005-AGNSam} in the Sloan Digital Sky Survey (SDSS). Out of 42\,435 galaxies, 1317 have an H$\alpha$ full width at half maximum larger than 1200\,km s$^{-1}$, and should be classified as Seyfert 1's, i.e.\ about 3\%. Since the classification is based on H$\alpha$, this includes intermediate types. Seyfert 2 galaxies are not so easily identified, and different selection criteria in the standard line-ratio diagrams \citep{VeilEtal-1987-SpeCla} lead to different fraction of Seyfert galaxies. Using the \citet{KewlEtal-2001-TheMod} theoretical separation between star-forming galaxies and AGN, \citet{HaoEtal-2005-AGNSam} find 3074 completely AGN-dominated narrow-line galaxies, i.e. about 7\%. Using the more inclusive criteria of \citet{KaufEtal-2003-HosGal}, which also selects AGN-starbursts composite galaxies, the figure increases up to 10\,700 galaxies, i.e. 25\%. The ratio between Seyfert 1 and Seyfert 2 galaxies is therefore between 0.12 and 0.43. We find here a significantly larger ratio of 1.13. 
Therefore a large fraction of Seyfert 2 galaxies and/or Seyfert 2-starbursts composite are invisible in hard X-rays. It has been observed by \citet{StefEtal-2003-ChaAct} that the AGN 2-8\,keV luminosity function was dominated by broad-line AGN at high X-ray luminosities and by narrow-line AGN at lower luminosities, contrarily to what is expected in the unified model of AGN \citep[e.g.,]{Anto-1993-UniMod}. As we sample mostly the high-luminosity population, this can explain the discrepancy. \citet{BargEtal-2005-CosEvo} analyzed in more detail the dependence of AGN type fraction on X-ray luminosity, and postulated that it could indicate that the covering fraction of the absorbing material decreases with luminosity, a scenario compatible with the receding-torus model used to account for a similar AGN-type dependence on radio luminosities \citep{Lawr-1991-RelFre}. The limited statistics prevent us from checking whether this effect can completely explain the fraction of broad-line AGN we observe here, but we point out that there could be an alternative explanation to our result if a significant fraction of the Seyfert 2 galaxies are absorbed with column densities in excess of $10^{25}$\,cm$^{-2}$, which would make them undetectable even in the hard X-rays.

Deep X-ray surveys \citep[e.g.][]{GiacEtal-2001-FirRes} showed the existence of apparently normal galaxies with strong X-ray emission, the so-called X-ray bright optically normal galaxies \citep[XBONG;][]{FioreEtal-2000-SpeInd}. In this survey, we detect one XBONG, NGC~4992. While these sources consist most probably of a mix of different kinds of objects \citep{GeorGeor-2005-XraBri}, the high luminosity of this source ($L_{\mathrm X}\sim 10^{43.5}$\,erg s$^{-1}$), makes that it is quite probably a true AGN.

\subsection{Intrinsic absorption}
\begin{figure}
\includegraphics[width=8.8cm]{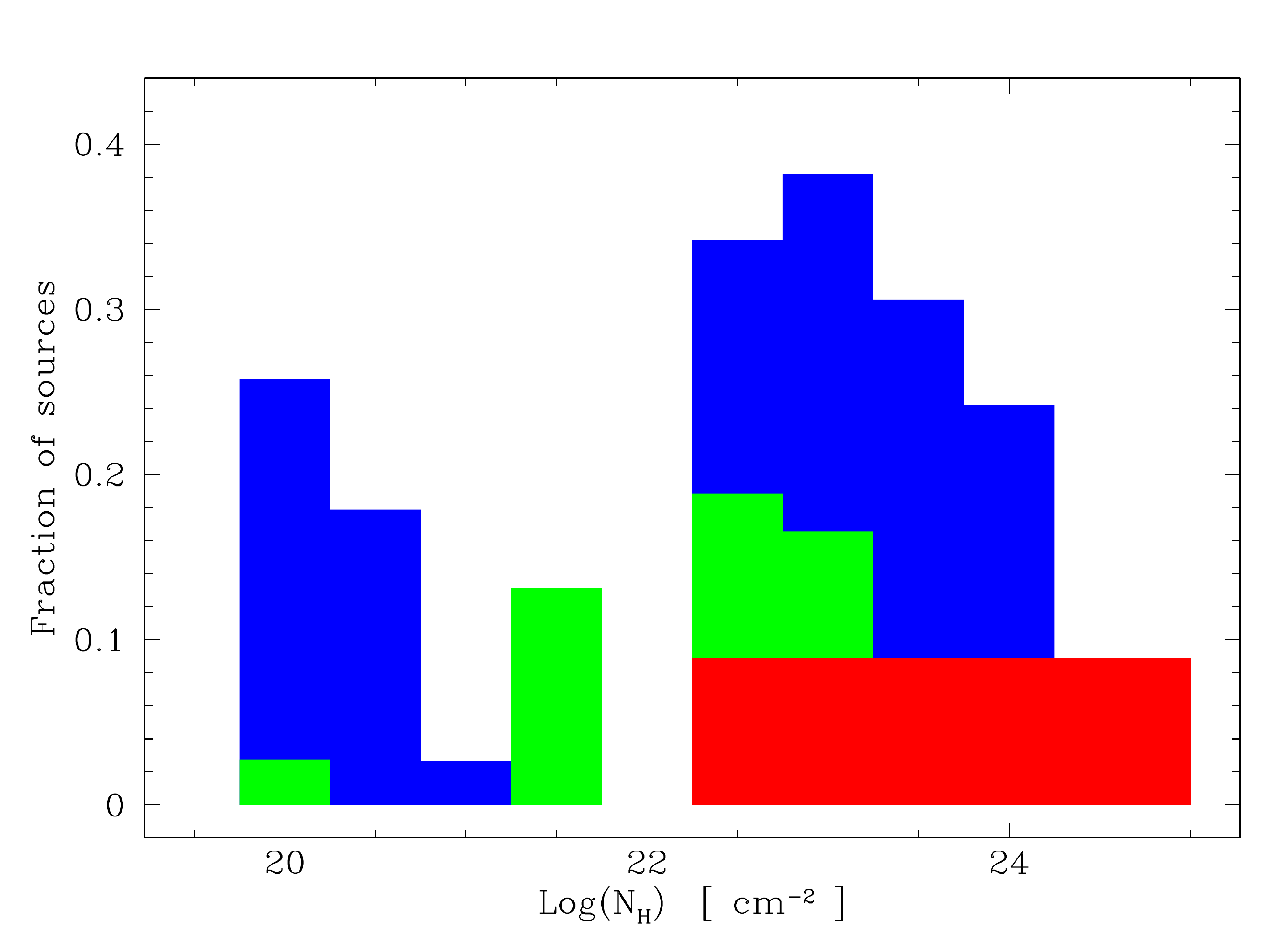}
\caption{\label{fig:nh}
Histogram of the \nh\ distribution for the sources from Table~\ref{tab:prop}. Sources are weighted by their probabilities. The blue histogram shows sources with measured intrinsic \nh's. The green histogram shows the sources whose \nh\ is derived from the presence of a counterpart in the RASS-BSC. The red histogram shows the sources without counterparts in the RASS-BSC, with log\,\nh\ evenly distributed between $22.25$ and $25$.}
\end{figure}
Figure~\ref{fig:nh} shows the distribution of the intrinsic hydrogen column density \nh\ for the 34 sources detected in the mosaic and weighted by their probabilities. The fraction of absorbed objects, i.e.\ those with \nh$>10^{22}$\,cm$^{-2}$ is found to be 70\%; if one discard the 11 sources without X-ray counterparts, this figure becomes 46\%, making it a stringent lower limit. None of the 23 sources with measured or estimated \nh\ are Compton-thick (\nh$>10^{24}$\,cm$^{-2}$). We therefore obtain an upper limit to the fraction of Compton-thick objects of 24\%. These figures are consistent with those found in previous hard X-ray surveys. In their modeling of the cosmic X-ray background, \citet{GillEtal-2007-SynCos} predict that, at the level of $10^{-11}$\,erg s$^{-1}$ cm$^{-2}$, the fraction of absorbed AGN (with \nh$>10^{22}$\,cm$^{-2}$) should be 65\% is perfectly compatible with our measurement. Their expected fraction of Compton-thick AGN (15\%) is also compatible with our upper limit, although the lack of detection of any true Compton-thick object makes our constraint rather weak.

There is a suggestion in Fig.~\ref{fig:nh} that there is a lack of objects with \nh\ about $10^{21-22}$\,cm$^{-2}$. The chance of observing such a drop by chance from a flat distribution in Log \nh\ is about 10\%, which is not very significant, but the \nh\ distribution of SWIFT/BAT AGN \citep{TuelEtal-2007-SwiBat} and in the previous INTEGRAL/IBIS survey of \citet{BeckEtal-2006-HarXra} are both compatible with a similar drop. If verified on a larger sample, this drop could result from a double origin of the absorption: one absorber with large \nh, as in the torus model; and one absorber with moderate \nh\ in a galactic-type distribution.

In the examination of local Seyfert 2 galaxies by \citet{GuaiEtal-2005-XraObs}, about 50\% were found to be Compton-thick, i.e. with absorbing columns in excess of $10^{24}$\,cm$^{-2}$.  For the sample of 20--60\,keV sources presented here, the median redshift (for the AGN with redshift information) is $z = 0.023$, and the median 20--60\,keV luminosity is $10^{43.5}$\,erg\,s$^{-1}$, i.e. we are probing a slightly more distant and higher luminosity population than that of the Guainazzi sample.

We find that all Seyfert 2 galaxies are absorbed (\nh$\ge 10^{22}$\,cm$^{-2}$), while none of the Seyfert 1 galaxies are. Among the intermediate types, 2 are absorbed and 2 are not. This confirms the relationship between the absence of broad lines and obscuration found in local Seyfert galaxies \citep{CappEtal-2006-XraSpe}, which is again a natural consequence of the geometrical unification model of Seyfert galaxies \citep{Anto-1993-UniMod}.

The largest intrinsic absorption is actually observed for the XBONG NGC~4992, with \nh$= 0.9\,10^{24}$\,cm$^{-2}$. This object seems similar to \object{CXOU J$031238.9-765134$}, which, according to \citet{ComaEtal-2002-HelII}, is a very strongly absorbed AGN. If Compton-thick, its hard X-ray luminosity should be about $10^{44}$\,erg s$^{-1}$, close to that of NGC~4992.
Using X-ray observations below 10\,keV \citet{GeorGeor-2005-XraBri} found that XBONGs are a mix of quite different objects. However, many of these objects, like those presenting thermal-like emission with a temperature on the order of 1\,keV, will not be detectable in the hard X-rays. Thus, it seems that hard X-ray selected XBONGs may to a large extent consist of a very absorbed population of AGN.

\begin{figure}
\includegraphics[width=8.8cm]{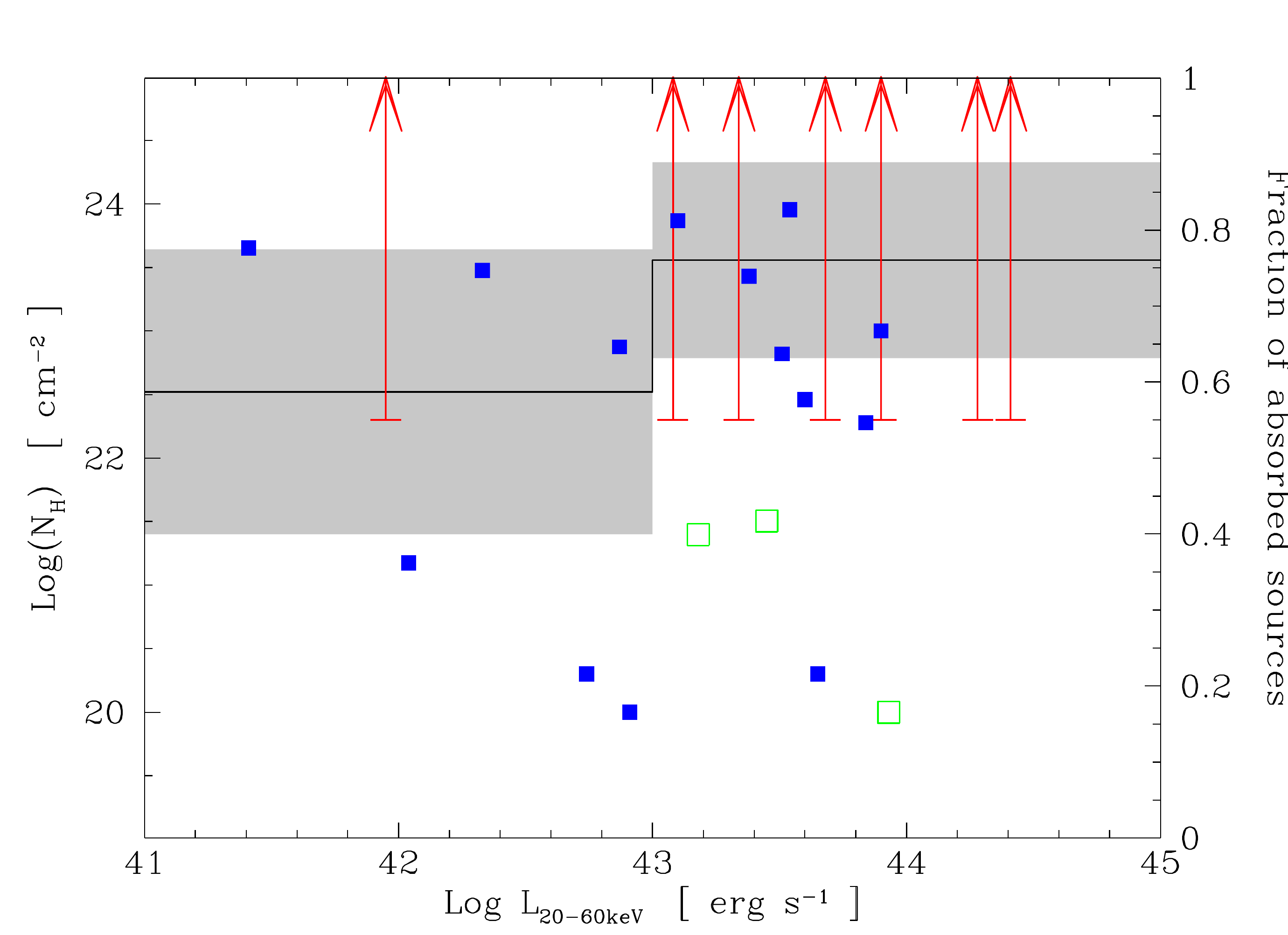}
\caption{\label{fig:lumnh}
\nh\ vs luminosity for all sources with significance $\sigma>5$, the Coma cluster and blazars having been excluded. Blue filled squares correspond to objects with actual \nh\ measurements. Green open squares correspond to objects whose \nh's have been determined using their fluxes in RASS-BSC. Red arrows are lower limits based on the absence of X-ray counterparts. The heavy black line shows the fraction of absorbed AGN (\nh$\ge 10^{22}$\,cm$^{-2}$) in two luminosity bins, $L<10^{43}$ and $L>10^{43}$\,erg s$^{-1}$. The gray area shows the 1$\sigma$ uncertainties on these fractions.}
\end{figure}

Figure~\ref{fig:lumnh} shows the relation between \nh\ and hard X-ray luminosity, as well as the fraction of absorbed sources in two luminosity bins, $L<10^{43}$ and $L>10^{43}$\,erg s$^{-1}$ respectively. A strong anticorrelation between the fraction of absorbed source and luminosity has been found in $2-10$\,keV surveys by \citet{UedaEtal-2003-CosEvo} and later confirmed by several groups \citep{LafrEtal-2005-HarXra,GeorEtal-2006-DeeCha,GillEtal-2007-SynCos}, although \citet{DwelPage-2006-DisAbs} question this anticorrelation on the basis of extensive simulations, arguing that the discrepancy might results from the combination cosmic variance (many results coming from the Chandra Deep Field South) and observation biases due to the limited bandwidth of Chandra compared to XMM-Newton. No correlation can be seen in our sample; the fraction of absorbed AGN is compatible with being constant around 70\%. Among the three previous hard X-ray surveys, only \citet{SazoEtal-2007-HarXra} may have found an anticorrelation with some credible significance. In any case, hard X-ray samples are still too small to exclude or confirm the anticorrelation claimed in the medium X-rays.

\subsection{Hard X-ray source counts}

\begin{figure*}
\includegraphics[angle=0,width=\textwidth]{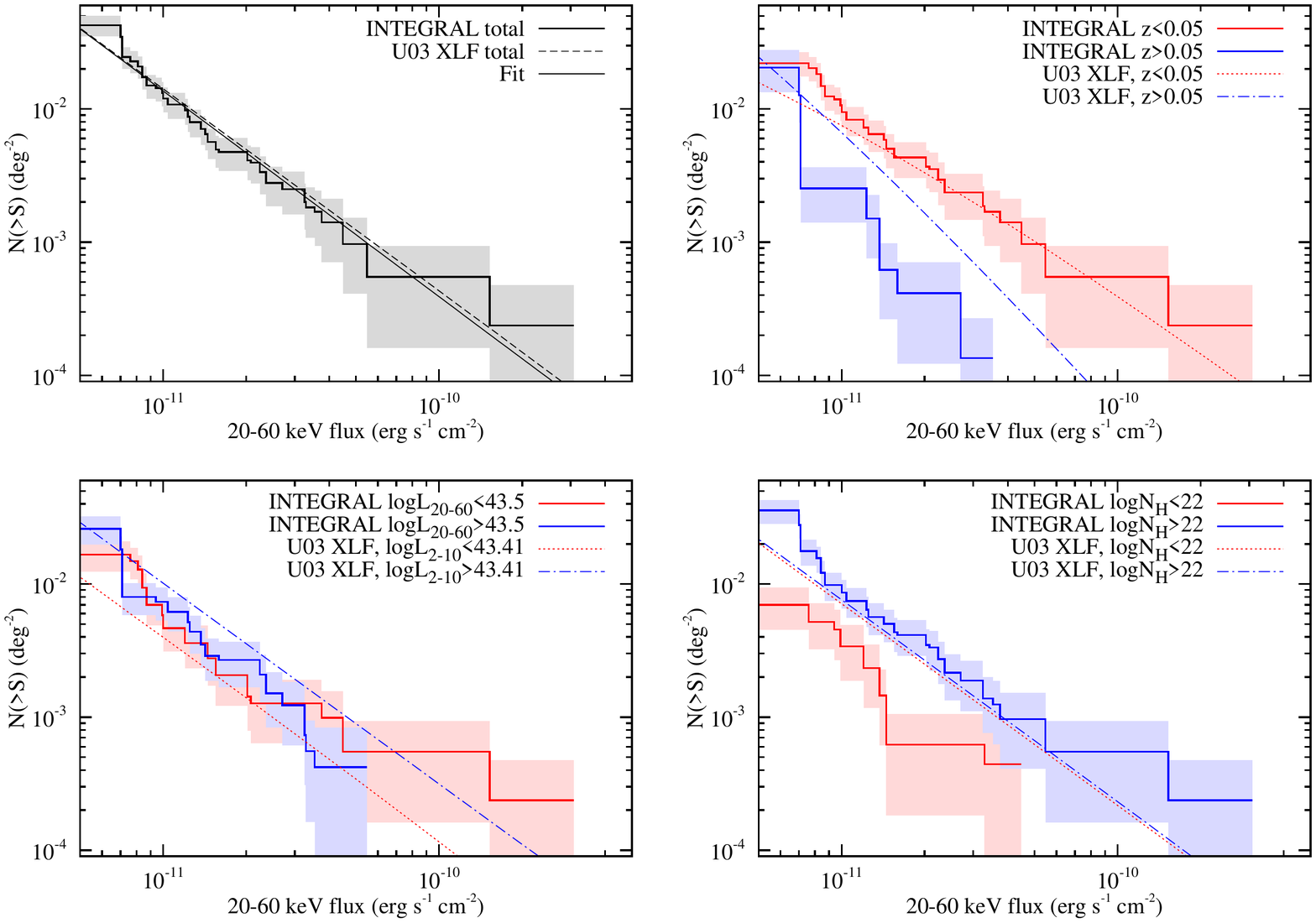}
\caption{\label{fig:models}
$\log N$--$\log S$ diagrams for the sources detected by INTEGRAL. In all panels the solid lines are from our work and the dashed or dotted line are model extrapolations from U03. Top left: $\log N$--$\log S$ for the full list of candidate sources with the 1$\sigma$ uncertainties. The solid line is the power-law best fit. Top right: $\log N$--$\log S$ in two redshift bins; red: $z < 0.05$; blue: $z>0.05$. Bottom left: $\log N$--$\log S$ in two luminosity bins; red: $L_{20-60\,\mathrm{keV}} < 10^{43.5}$; blue: $L_{20-60\,\mathrm{keV}} > 10^{43.5}$.
Bottom right: $\log N$--$\log S$ in two \nh\ bins; red: \nh$< 10^{22}$\,cm$^{-2}$; blue: \nh$> 10^{22}$\,cm$^{-2}$.}
\end{figure*}

Figure~\ref{fig:models} top left shows the $\log N$--$\log S$ diagram of the sources detected in the mosaic. To calculate this diagram we used all the candidate sources weighted by their probabilities of being real.
3C~273 and Coma cluster were however discarded from the $\log N$--$\log S$ diagram, because they were targeted by the observations, and their presence in the mosaic is not due to chance; in addition, Coma cluster is a cluster of galaxies and its high-energy emission is most probably not due to the presence of an AGN \citep{EckeEtal-2007-SouWes}. The faintest fluxes are approximately 0.6\,10$^{-11}$\,erg s$^{-1}$ cm$^{-2}$. At this flux limit, the source surface density is $0.033\pm 0.010$\,deg$^{-2}$. At the level of $10^{-11}$\,erg s$^{-1}$ cm$^{-2}$, source surface density reaches $0.014\pm 0.0025$\,deg$^{-2}$. The best-fit relationship is $\log N=-1.55\pm 0.13\,\log S -1.86\pm 0.07$, $S$ being the 20--60\,keV flux in units of $10^{-11}$\,erg s$^{-1}$ cm$^{-2}$. The slope is in quite good agreement with the expected $-3/2$ slope.

We further point out that, at the sensitivity limit of this survey, INTEGRAL resolves about 2.5\% of the cosmic X-ray background \citep{GrubEtal-1999-SpeDif,ChurEtal-2007-IntObs}; this fraction is about twice that found in \citet{BeckEtal-2006-HarXra}. A turnover in the $\log N$--$\log S$ relationship must be present at a flux between 10$^{-12}$ and 10$^{-14}$\,erg s$^{-1}$ cm$^{-2}$, similar to what is observed in the medium X-rays.

\subsubsection{Comparison with other surveys}

At the level of $2\,10^{-11}$\,erg s$^{-1}$ cm$^{-2}$ between 20 and 60\,keV (which is a flux limit reached by all surveys) \citet{BeckEtal-2006-HarXra} find a source density of $4.3\,10^{-3}$\,deg$^{-2}$, very close to our density of $4.7\,10^{-3}$\,deg$^{-2}$. \citet{TuelEtal-2007-SwiBat}, using SWIFT/BAT all-sky survey, find indeed a density of $2.2\,10^{-3}$\,deg$^{-2}$, lower than the INTEGRAL density by a factor 2. The authors point out that differences in Crab calibration might be the cause of this discrepancy, but the 15\% difference in the Crab flux falls short of a factor 2 in source density. We checked that the assumption on the exact AGN spectral shape, for instance absorption, has a negligible effect. Thus the discrepancy between INTEGRAL and SWIFT/BAT counts remains mostly unexplained.

The 20--60\,keV source counts have a very similar slope and normalization to the bright end of the 2--10\,keV source counts. From a compilation of 2--10\,keV surveys, \citet{CarrEtal-2007-XmmNew} find a best fitting slope of -1.58 and normalization of 485.3\,deg$^{-2}$ at $1.17\times 10^{-14}$\,erg\,s$^{-1}$\,cm$^{-2}$, equivalent to 0.0113\,deg$^{-2}$ at $10^{-11}$\,erg\,s$^{-1}$\,cm$^{-2}$. Likewise, the 2--10\,keV source counts presented by \citet{MoreEtal-2003-ResFra} have a slope of -1.57 with a slightly lower normalization of 0.0082\,deg$^{-2}$ at $S_{2-10keV} = 10^{-11})$\,erg\,s$^{-1}$\,cm$^{-2}$. It should be pointed out that the 2--10\,keV source counts are only well defined at very faint fluxes (compared to hard X-ray surveys), and so a small uncertainty on the slope of the 2--10\,keV source counts can make a fairly large uncertainty when extrapolating up to $\sim 10^{-11}$\,erg cm$^{-1}$ s$^{-1}$.

If we are seeing the same source population at 2--10 and at 20--60\,keV, then to reconcile the 2--10\,keV source count normalizations of \citet{MoreEtal-2003-ResFra} and \citet{CarrEtal-2007-XmmNew} with that seen here at 20--60\,keV then the mean source flux ratio, $S_{20-60\,{\mathrm{keV}}}/S_{2-10\,{\mathrm{keV}}}$, must be somewhere between 1 and 1.5.  If we make the assumption that the average source spectrum is a power-law we require a mean slope of $\Gamma \sim$ 1.5--1.8 to make the source count normalizations agree with each other.  Provided that sources without measured redshift do not lie at high redshift (and hence they don't need any significant K-correction), the exponential cut-off at about $200$\,keV \citep{PeroEtal-2002-ComRef} in the average spectrum does not affect significantly the average spectral slope. This average slope of 1.5--1.8 is consistent with those found in AGN surveys at lower X-ray energies \citep{PicoEtal-2003-XmmNew,MateEtal-2005-XmmNew}. While this average slope, which is significantly harder than the typical $\Gamma=1.9$ intrinsic emission, implies the presence of absorbed sources, it suggests that at the 20--60\,keV flux level probed by our survey, there is no significant new population of Compton-thick sources that are not detected by surveys at lower energies, as the effective power-law slope between these two bands would be much harder.

\subsubsection{Comparison with models of the 2--10\,keV AGN population}
In Fig.~\ref{fig:models} top left we compare our source counts to the predictions of the 2--10\,keV population model of \citet{UedaEtal-2003-CosEvo} (hereafter U03).  We predict the 20--60\,keV source counts by integrating the U03 model AGN population over the $0 < z < 2$, $10^{41} < L_{2-10\,\mathrm{keV}} < 10^{47}$\,erg s$^{-1}$ range. The conversion from rest-frame intrinsic (i.e. before absorption) 2--10\,keV luminosity to observed frame 20--60\,keV flux is made using the same
spectral model adopted by U03, namely a power-law spectrum with $\Gamma = 1.9$, a cut-off rest-frame energy of 500\,keV, and a reflection component from cold material. With this spectral model, $S_{20-60\,\mathrm{keV}} = 1.24 \times L_{2-10\,\mathrm{keV}}/4 \pi d^{2}_{\mathrm L}$ at redshift $\sim 0$ ($d_{\mathrm L}$ is the luminosity distance in cm). We use here the U03 model which includes a
mix of unabsorbed and Compton-thin sources, but no Compton-thick sources; the Compton-thick population has indeed not been measured and has been treated somewhat arbitrarily by adding a number of these objects equivalent to that of the Compton-thin ones. We have therefore assumed that absorption effects on this population in the 20--60\,keV band are negligible.

Under these assumptions, we can see that the U03 model provides a good match to both the slope and normalization of the total 20--60\,keV source counts (see Fig.~\ref{fig:models} top left),
leaving little room for a significant additional population of moderately Compton-thick sources. We examine the population in more detail by splitting the sample into low ($z<0.05$) and high redshift sources, low ($L_{20-60\,\mathrm{keV}} < 10^{43.5}$\,erg s$^{-1}$) and high luminosity sources, and low (\nh$< 10^{22}$\,cm$^{-2}$) and high absorption sources. In Fig.~\ref{fig:models} top right we show the source counts and model predictions for redshifts either below, or above 0.05. Redshifts are unknown for 6 sources. On the basis of their optical magnitudes and alternative possible identifications we assume that all these sources, with the exception of IGR J12042$-$0756 and IGR J12060+3818, lie at $z > 0.05$ (and hence have 20--60\,keV luminosities in excess of $10^{43.5}$\,erg s$^{-1}$). The observed counts and model predictions agree reasonably well given the relatively small numbers of observed sources. Above $S_{20-60\,\mathrm{keV}} \sim 10^{-11}$\,erg cm$^{-2}$ s$^{-1}$, the U03 model overpredicts the number of observed sources with $z>0.05$. Irrespective of our redshift assumptions, the 20--60\,keV integral source counts are dominated at all fluxes by low-redshift objects.

In Fig.~\ref{fig:models} bottom left we show the source counts and model predictions for observed sources above and below a luminosity of $L_{20-60\,\mathrm{keV}} = 10^{43.5}$\,erg s$^{-1}$. For our given spectral model this corresponds to $L_{2-10\,\mathrm{keV}} = 10^{43.41}$\,erg s$^{-1}$ (as
described above). The observed counts and model predictions are roughly in agreement, given the small number of sources. Note however that the U03 model predicts roughly twice as many sources with high luminosity as with low luminosity, but that approximately equal numbers of sources are observed above and below $L_{20-60\,\mathrm{keV}} = 10^{43.5}$\,erg s$^{-1}$.

Absorbing column estimates or lower limits are available for our entire sample and so, in Fig.~\ref{fig:models} bottom right, we show the source counts and model predictions separated into sources with \nh\ either greater than, or less than $10^{22}$\,cm$^2$. Here it is clear that the U03 model is a poor predictor of the observed source counts. We see that the more absorbed sources constitute at least 2/3 of the total 20--60\,keV source counts over the flux range of the sample, whereas the UO3 model predicts equal number of absorbed and non-absorbed sources over the luminosity and redshift range probed by the INTEGRAL observations. Taken at face value, it means that the observed and model $N_H$ distributions differ significantly. A number of our \nh\ measurements use indirect method based on the presence of the ROSAT flux or non-detection. A follow-up with more
sensitive medium energy X-ray observations is therefore needed to confirm this result.

We have also investigated the dependence of the predicted source counts on the exact choice of spectral model. We find that the predicted source counts are relatively insensitive to the choice of cut-off energy, which is to be expected because the source counts are dominated by low redshift objects. The overall normalization of the predicted source counts is rather dependent on the spectral slope and the size of the reflection component. A harder spectral slope or stronger reflection component increases the total 20--60\,keV counts predicted by the U03 model. However, changing the spectral model has very little effect on the predicted ratios of high to low redshift sources, high to low luminosity sources or high to low absorption sources.

\subsection{Hard X-ray luminosity function of local AGN}
\begin{figure*}
\includegraphics[width=8.8cm]{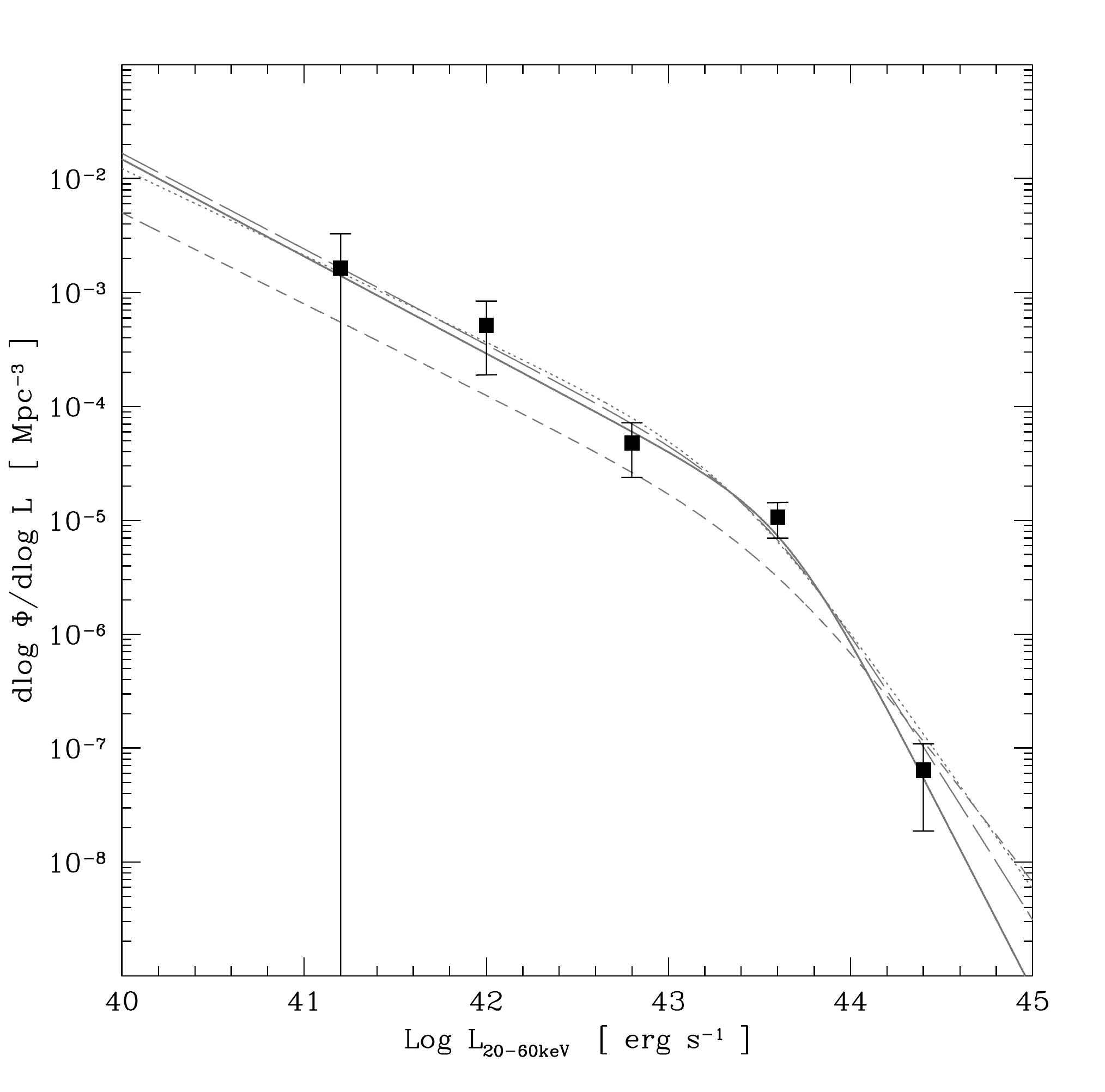}\hfill\includegraphics[width=8.8cm]{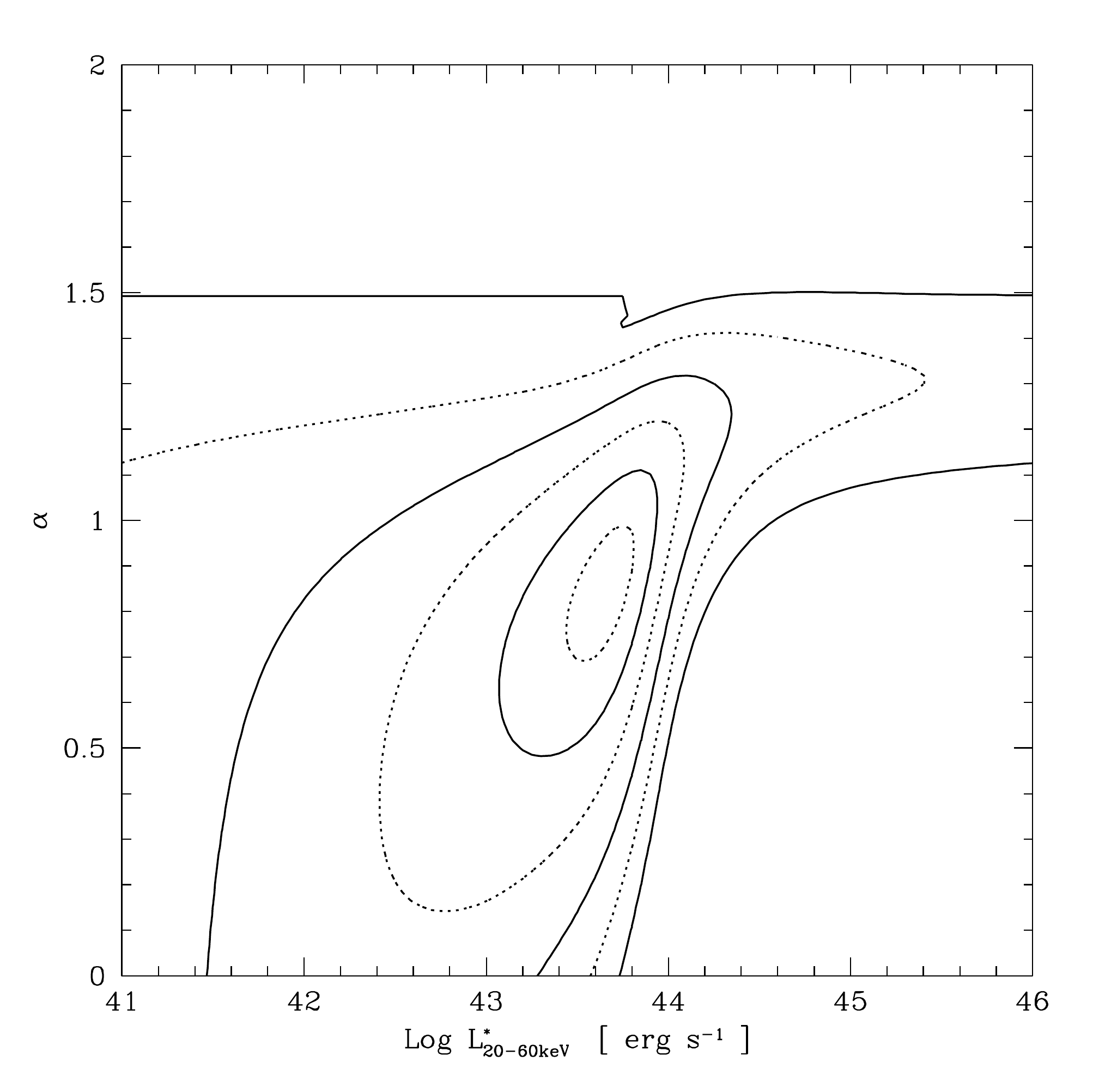}%
\caption{\label{fig:lf} Left: AGN luminosity function in the 20-60\,keV range. The points have been calculated using the $V/V_{\mathrm{max}}$ estimator. The heavy solid line is the $\Phi(L)$ distribution from Eq.~(\ref{eq:lf}) with $L^*$, $\alpha$ and $\beta$ parameters determined using a maximum-likelihood method and $\Phi^*$ obtained by fitting the $V/V_{\mathrm{max}}$ points. Short-dashed line is the LF from \citet{BeckEtal-2006-HarXra}; dotted line is that from \citet{SazoEtal-2007-HarXra}; long-dashed line is that from \citet{TuelEtal-2007-SwiBat}. Right: Uncertainty contours in the $L^*$--$\alpha$ plane. The solid and dotted lines mark the 1,2,3 and 0.5,1.5,2.5 $\sigma$ thresholds respectively.}
\end{figure*}
We use the AGN firmly detected in this work (i.e. marked with a star in Table~\ref{tab:sources} to determine the AGN luminosity function (LF) in the 20--60\,keV energy domain. However, because of the shallow depth of INTEGRAL surveys, the volume adequately sampled is quite small, making the high-luminosity tail of the LF, where objects are very rare, impossible to determine. We therefore complement our sample with the sample of \citet{BeckEtal-2006-HarXra} to study the part of the LF with $L_{20-60}>10^{44}$\,erg s$^{-1}$, using a clear cut in luminosity to avoid source duplication in these overlapping surveys. We convert the luminosities from the 20--40\,keV to 20--60\,keV domain using the flux conversion from Sect.~\ref{sec:prop}. We chose to study the redshift bin $0\le z\le 0.05$ only to study the really local population and to minimize any effect of evolution; this has also the benefit of discarding the three blazars. We end up with 19 AGN from our sample and 2 from \citet{BeckEtal-2006-HarXra}.

Fig.~\ref{fig:lf} left shows the luminosity function obtained from these samples. Because of the small number of sources, we use a parametric method to derive the LF. We assume the standard AGN luminosity function:
\begin{equation}
\frac{{\mathrm d}\Phi(L)}{{\mathrm d}\log L}=\frac{\Phi^*}{(L/L^*)^{-\alpha}+(L/L^*)^{-\beta}},
\label{eq:lf}
\end{equation}
which describes a broken power-law, changing from index $\alpha$ to index $\beta$ at characteristic luminosity $L^*$. From the area-detection threshold relationship, we build the probability function $P_{\mathrm{det}}(L)$, which gives the probability that a source with a luminosity $L$ and a redshift $z<0.05$ is detected in this survey. Given a luminosity function $\Phi(L)$, we can derive the probability distribution of $L$ in our survey. The parameters $L^*$, $\alpha$ and $\beta$ parameters are then determined using a maximum-likelihood (ML) test, based on the idea proposed by \citet{SandEtal-1979-VelFie}, modified to take into account our sensitivity map. As this method is independent of the normalization $\Phi^*$, we also estimate the LF using the standard non-parametric $V/V_{\mathrm{max}}$ estimator. $\Phi^*$ is then obtained by fitting the $V/V_{\mathrm{max}}$ LF estimates with the $\Phi(L)$ distribution from Eq.~(\ref{eq:lf}), letting only $\Phi^*$ free and fixing the other parameters to the ML values.

To determine the uncertainties on the estimated parameters, we use the standard Gaussian approximation of the maximum-likelihood peak, i.e.\ the $n$-sigma confidence intervals for $\alpha$ are determined using the relationship: $\Delta{\cal L}(L^*,\alpha,\beta)=\exp(-\frac{n^2}{2})$, where $\cal L$ is the likelihood and $\Delta{\cal L}$ is the ratio between the likelihood for a given set of parameters and the maximum likelihood.
The $(L^*,\alpha)$ uncertainty contours are shown in Fig.~\ref{fig:lf} right. Confidence interval on $\Phi^*$ is obtained by drawing sets of parameters $(L^*,\alpha,\beta)$ according to their likelihood and by fitting $\Phi^*$ on the $V/V_{\mathrm{max}}$ LF estimates. Finally, we obtain:
\begin{equation}
 \begin{array}{ccl}
  \log L^*&=&43.66\,^{+0.28}_{-0.60}{\mathrm {~erg~s}}^{-1}\\[1mm]
  \alpha&=&\z 0.85\,^{+0.26}_{-0.38}\\[1mm]
  \beta&=&\z 3.12\,^{+1.47}_{-1.02}\\[1mm]
  \Phi^*&=&\z 1.12\,^{+5.04}_{-0.77}~10^{-5}{\mathrm {~Mpc}}^{-3}\\
\end{array}
\end{equation}
 Fig.~\ref{fig:lf} right shows, as an example, the uncertainty contours in the $L^*-\alpha$ plane. $\Phi^*$ is badly constrained, because it is very strongly correlated to $L^*$; therefore the LF normalization is known with a much better accuracy than the uncertainty on $\Phi^*$ suggests.

Figure~\ref{fig:lf} left also shows LFs determined from other surveys in similar energy ranges \citep{BeckEtal-2006-HarXra,SazoEtal-2007-HarXra,TuelEtal-2007-SwiBat}\footnote{We note that in the submitted version of the \citet{TuelEtal-2007-SwiBat} paper their LF is plotted with a wrong normalization, lower than the value they quote}. The LFs translated to the 20--60\,keV range are very consistent with each other, except for the \citet{BeckEtal-2006-HarXra} LF, which shows a lower normalization by a factor about 2. The $L^*$, $\alpha$ and $\beta$ parameters are consistent with each other in all studies. While the slopes $\alpha$ and $\beta$ obtained in the 2--10\,keV domain \citep[e.g.,][]{LafrEtal-2005-HarXra} are quite comparable to ours, the characteristic luminosity $L^*$ is a factor about 4 higher than that observed here. As the conversion factor $S_{20-60\,{\mathrm{keV}}}/S_{2-10\,{\mathrm{keV}}}$ obtained by comparing the Log\,$N$--Log\,$S$ diagrams was found to be between 1 to 1.5, this gives a difference by a factor about 5 in energy-corrected $L^*$. This is formally compatible with our study, but we note that all hard X-ray studies find $L^*$ values similar to ours, making the case for a statistical fluctuation rather weak. As \citet{LafrEtal-2005-HarXra} use an evolution model over four redshift bins, the lowest one being $0\le z\le 0.5$, it might be that the $z=0$ LF suffers from an extrapolation effect. It is however in line with the comparison of the source count diagrams, where it appears that medium X-ray surveys see a less absorbed (and hence more luminous) AGN population.

The spatial density of AGN with luminosity $L_{\mathrm{20-60\,keV}}>10^{41}$\,erg s$^{-1}$ is:
\begin{equation}
  \Phi(L>10^{41}) = \int_{41}^\infty \frac{\mathrm{d}\Phi}{\mathrm{d}\log L}\,\mathrm{d}\log L=\z1.03\,^{+0.98}_{-0.64}~10^{-3}{\mathrm {~Mpc}}^{-3}
\end{equation}
This value is perfectly compatible with \citet{SazoEtal-2007-HarXra} ($1.18~10^{-3}$\,Mpc$^{-3}$) and with \citet{TuelEtal-2007-SwiBat} ($1.23~10^{-3}$\,Mpc$^{-3}$), while \citet{BeckEtal-2006-HarXra} gives a lower value of $0.42~10^{-3}$\,Mpc$^{-3}$.

The luminosity density integrated over $L_{20-60\,\mathrm{keV}}=10^{41}$\,erg s$^{-1}$ is:
\begin{equation}
 \begin{array}{ccl}
  W(L>10^{41})&=&\int_{41}^\infty L\frac{\mathrm{d}\phi}{\mathrm{d}\log L}\,\mathrm{d}\log L =\\[2mm]
  &=&\z0.90\,^{+0.19}_{-0.25}~10^{39}{\mathrm {erg~s}}^{-1}\mathrm{{~Mpc}}^{-3}
\end{array}
\end{equation}
Again, this is compatible with \citet{SazoEtal-2007-HarXra} and \citet{TuelEtal-2007-SwiBat} ($W(L>10^{41})=1.02~10^{39}$ and $1.03~10^{39}$ erg s$^{-1}$ Mpc$^{-3}$ respectively), and about twice the luminosity density ($0.41~10^{38}$ erg s$^{-1}$ Mpc$^{-3}$ found in \citet{BeckEtal-2006-HarXra}, taking the flux correction into account.

\section{Summary and conclusion}

We presented an extragalactic survey of about 2500\,deg$^{2}$ with INTEGRAL/IBIS. With a flux limit of approximately $0.6~10^{-11}$\,erg cm$^{-2}$ s$^{-1}$ in the 20--60\,keV range, this is the deepest hard X-ray extragalactic survey to date. We detected 34 candidate sources, for which we searched for counterparts in other wavelength domains. We find that the ratio of detected Seyfert 1 vs Seyfert 2 galaxies is larger than in optically selected samples. This suggests that current hard X-ray surveys are biased against Seyfert 2 galaxies, either because these surveys are limited to high-luminosity objects or because some sources are extremely absorbed sources, with \nh\ above $10^{25}$\,cm$^{-2}$.

We studied the distribution of absorption in our objects. While we do not find any Compton-thick object, using object without \nh\ determination and without counterpart in the ROSAT-BSC catalog, we  find that at most 24\% of the sources are Compton-thick. The fraction of absorbed object (\nh$>10^{22}$\,cm$^{-2}$) is between 46 and 70\%. Although the small number of objects makes the result not significant, there's is a hint that the distribution could be bimodal, with absorption originating either from the host galaxy or from a thick structure with non-unity covering factor, like a dust torus. We do not find any relationship between \nh\ and luminosity, but, because of the small number of objects, we cannot exclude it. We note however that none of the current hard X-ray surveys find such relationship with any credible significance.

The Log\,$N$--Log\,$S$ diagram of our candidate sources has been found to reach 0.013\,deg$^{-1}$ at the level of $10^{-11}$\,erg cm$^{-2}$ s$^{-1}$. We resolved approximately 2.5\% of the cosmic X-ray background. Comparison with the 2--10\,keV domain shows that the Log\,$N$--Log\,$S$ diagrams are compatible provided the average effective power-law index between the 2--10\,keV and the 20--60\,keV domains is $\Gamma\sim 1.5-1.8$, consistent with the 2-10\,keV slope found in medium X-ray surveys. This shows that there isn't any large population of bright Compton-thick objects missed in the 2--10\,keV surveys and appearing in the hard X-rays.

We performed a detailed comparison of our source counts with those predicted from the population model of \citet{UedaEtal-2003-CosEvo}. Source counts are in general consistent, the biggest difference being in the Log\,$N$--Log\,$S$ diagram of sources separated in two \nh\ bins, \nh$<10^{22}$ and \nh$>10^{22}$\,cm$^{-2}$.  As a large fraction of our \nh's were estimated based on ROSAT observations, our \nh\ distribution must be confirmed in follow-up observations in the medium X-rays.

We present a truly local hard X-ray luminosity function ($z<0.05$). We find a LF quite compatible with the latest all-sky INTEGRAL and SWIFT surveys. The characteristic luminosity $L^*$ is however a factor about 5 lower than that in the 2--10\,keV range. This discrepancy is seen in all hard X-ray studies of the local LF, and would imply, similarly to what is seen in source count diagrams, that the 2--10\,keV population is less absorbed than that seen in the hard X-rays.

The population of Compton-thick objects that is expected from models of the cosmic X-ray background and which should appear in the hard X-rays is still elusive. While the fraction of objects which are possibly Compton-thick is compatible with the models, the lack of clearly Compton-thick objects in our sample and the good match with the source counts from models derived from medium X-ray observations make that the case for the existence of such population is rather weak. Follow-up observations of our sources without adequate \nh\ measurements are in progress, and may solve the puzzle. It seems however quite probable that we shall end-up with conflicting \nh\ distributions, which will have to be explained.

While still suffering from low sensitivities compared to their modern counterparts working in the soft and medium X-rays, INTEGRAL and SWIFT are the only instruments currently available to perform surveys above 15\,keV. The importance of this energy domain, which is unaffected by obscuration below $\sim 10^{25}$\,cm$^{-2}$, is such that efforts to build statistically representative samples must be pursued.

\begin{acknowledgements}
This work is based on observations with INTEGRAL, an ESA project with instruments and science data centre funded by ESA member states (especially the PI countries: Denmark, France, Germany, Italy, Switzerland, Spain), Czech Republic and Poland and with the participation of Russia and the USA.

This research has made use of the SIMBAD database,
operated at CDS, Strasbourg, France.

This research has made use of the NASA/IPAC Extragalactic Database (NED) which is operated by the Jet Propulsion Laboratory, California Institute of Technology, under contract with the National Aeronautics and Space Administration.

This research has made use of data obtained through the High Energy Astrophysics Science Archive Research Center (HEASARC) Online Service, provided by the NASA/Goddard Space Flight Center.
\end{acknowledgements}

\bibliographystyle{apj}
\bibliography{biblio}

\end{document}